\documentclass[usenatbib]{mn2e}

\usepackage{graphicx}
\usepackage[usenames]{color}
\usepackage{amssymb}
\usepackage{appendix}

\newcommand{\zz}{z}
\newcommand{\Mpch}{h^{-1} \mbox{Mpc}}
\newcommand{\kpch}{h^{-1} \mbox{kpc}}
\newcommand{\pdm}{m_{\rm DM}}
\newcommand{\pgas}{m_{\rm gas}}
\newcommand{\fescionism}{f_{\rm esc,ion,ISM}}
\newcommand{\fesclyaism}{f_{\rm esc,Ly\alpha,ISM}}
\newcommand{\fesclyaigm}{f_{\rm esc,Ly\alpha,IGM}}
\newcommand{\msun}{\mbox{M}_{\odot}}
\newcommand{\Mdm}{M_{\rm DM}}
\newcommand{\Mgas}{M_{\rm gas}}
\newcommand{\Mstar}{M_{*}}
\newcommand{\Zgas}{Z}
\newcommand{\Zcrit}{Z_{\rm crit}}
\newcommand{\Zsun}{\mbox{Z}_{\odot}}
\newcommand{\Kelv}{\rm K}
\newcommand{\hpar}{h}
\newcommand{\ionfrac}{\mbox{x}_{\rm ion}}

\begin{document}

\date{Accepted 2012 May 23. Received 2012 May 23; in original form 2012 April 10}

\pagerange{\pageref{firstpage}--\pageref{lastpage}} \pubyear{2012}

\title[Effect of IGM on LAE detections during EoR]{Effect of intergalactic medium on the observability of Ly$\alpha$ emitters during cosmic reionization}
\author[A. Jeeson-Daniel et al.]
{Akila Jeeson-Daniel,$^{1}$\thanks{E-mail: akila@mpa-garching.mpg.de} Benedetta Ciardi,$^{1}$ Umberto Maio,$^{2}$ Marco Pierleoni,$^{1}$ \and Mark Dijkstra$^{1}$ and  Antonella Maselli$^{3}$\\ 
$^1$Max Planck Institute for Astrophysics, Karl-Schwarzschild Stra\ss{}e 1, 85741 Garching, Germany\\
$^2$Max Plank Institute for Extraterrestrial Physics, Gissenbachstra\ss{}e 1, 85748 Garching, Germany\\
$^3$EVENT Lab for Neuroscience and Technology, Universitat de Barcelona, 
    Passeig de la Vall d'Hebron 171, 08035 Barcelona, Spain}

\maketitle

\label{firstpage}

\begin{abstract}
We perform a systematic study of how the inhomogeneities in the intergalactic medium (IGM) affect the observability of Ly$\alpha$ emitters (LAEs) around the epoch of reionization. We focus on the IGM close to the galaxies as the detailed ionization distribution and velocity fields of this region could significantly influence the scattering of Ly$\alpha$ photons off neutral H atoms as they traverse the IGM after escaping from the galaxy. We simulate the surface brightness (SB) maps and spectra of more than 100 LAEs at $\zz=7.7$ as seen by an observer at $\zz=0$. To achieve this, we extract the source properties of galaxies and their surrounding IGM from cosmological simulations of box sizes $5-30\;\Mpch$ and then follow the coupled radiative transfer of ionizing and Ly$\alpha$ radiation through the IGM using CRASH$\alpha$. We find that the simulated SB profiles are extended and their detailed structure is affected by inhomogeneities in the IGM, especially at high neutral fractions. The detectability of LAEs and the fraction of the flux observed depend heavily on the shape of the SB profile and the SB threshold ($SB_{\rm th}$) of the observational campaign. Only ultradeep observations (e.g. $SB_{\rm th} \sim 10^{-23}$ erg~s$^{-1}$~cm$^{-2}$~arcsec$^{-2}$) would be able to obtain the true underlying mass-luminosity relation and luminosity functions of LAEs. The details of our results depend on whether Ly$\alpha$ photons are significantly shifted in the galaxy to longer wavelengths, the mean ionization fraction in the IGM and the clustering of ionizing sources. These effects can lead to an easier escape of Ly$\alpha$ photons with less scattering in the IGM and a concentrated SB profile, similar to the one of a point source. Finally, we show that the SB profiles are steeper at high ionization fraction for the same LAE sample which can potentially be observed from the stacked profile of a large number of LAEs.
\end{abstract}


\section{Introduction}

\cite{partridge67} predicted that high redshift galaxies would emit copious amounts of Ly$\alpha$ photons ($\sim 10$ per cent of the galaxy luminosity) due to the reprocessing of ionizing photons from the young massive stars by the interstellar medium (ISM). This stimulated a large number of studies aimed at observing the high-redshift galaxies which mostly lead to non-detections \citep[e.g.][and the references there in]{davis74,partridge74,djorgovski92}. Models were made to reason the null detections such as absorption of Ly$\alpha$ radiation by dust \citep[e.g.][]{spitzer78,meier81,hartmann88, charlot93} and lower number of massive 
stars due to ageing of stellar populations \citep{valls93}.

The first detections of high redshift Ly$\alpha$ emitting galaxies (which we also refer to as `Ly$\alpha$ Emitters', or LAEs) were reported in the mid-1990s \citep[e.g.][]{hu96}. Since then a large number of LAEs\footnote{The term `LAE' often refers to galaxies that have been selected on having a `strong' Ly$\alpha$ emission line. In this paper we use the term LAE for all Ly$\alpha$ emitting galaxies. Our classification thus includes ultraviolet (UV) selected galaxies for which follow-up spectroscopy revealed the presence of a strong Ly$\alpha$ emission line.} have been observed using both narrow band photometry and spectroscopy up to $\zz \lesssim 7$ \citep[e.g.][]{rhoads01,dawson04, rhoads04, taniguchi05, iye06, nilsson07, nilsson11, ouchi09, ouchi10,   wang09, guaita10, lehnert10, wolf10, cassata11,gronwall10,kashikawa11,pentericci11,ono11,schenker11}. In addition to being one of the main methods of detecting high-redshift galaxies to study the evolution of galaxy properties with redshift, LAEs can also be used to probe the epoch of reionization (EoR) \citep[e.g.][]{miralda98,malhotra04,mcquinn07} and to constrain dark energy properties using baryonic acoustic oscillations in the LAE clustering signal \citep[e.g.][]{eisenstein05,hill08}. Since this paper is motivated by the use of LAEs for the detection of EoR, we will discuss this in more detail.

Due to its resonant nature, the Ly$\alpha$ line is scattered by even very small amounts of neutral hydrogen along its path. The optical depth $\tau$ due to scattering away from the line-of-sight to the observer (at $\zz = 0$) for a LAE at redshift $\zz_{s}$ is given by
\begin{equation}
\tau \simeq 6.02 \times 10^{5}\;x_{\rm HI} \;\left(\frac{1+\zz_{s}}{10}\right)^{3/2},
\label{eqn:tau1}
\end{equation}
for photons entering the intergalactic medium (IGM) with wavelength $\lambda \lesssim \lambda_{\alpha} = 1215.67$~$\rm \AA$ \citep[Ly$\alpha$ rest wavelength; e.g.][]{gunn65,barkana01} and by
\begin{equation}
\tau=2.9\;x_{\rm HI} \left(\frac{\Delta v}{\rm 600\;km\;s^{-1}}\right)^{-1}\;\left(\frac{1+\zz_{s}}{10}\right)^{3/2},
\label{eqn:tau2}
\end{equation}
for photons entering the IGM with $\lambda = \lambda_{\alpha} \left(1+\Delta v/c\right)$ for an equivalent velocity shift of $\Delta v$ \citep[e.g.][]{miralda98,dijkstra10}. Here, $x_{\rm HI}$ is the neutral fraction of the IGM, $c$ is the speed of light in vacuum and the cosmology used to derive the above equations and for the rest of the work in this paper is the standard model - $\Omega_{0,m}$=0.3, $\Omega_{0,\Lambda}$=0.7, $\Omega_{0,b}$=0.04, $\hpar$=0.7, $n$=1 and $\sigma_{8}$=0.9. The effect of scattering on the shape of the intrinsic spectrum can be used to estimate the ionization fraction of the IGM around the source and thus study the EoR. 

LAEs luminosity functions (LFs), number density and clustering are the main methods used to study the EoR history. It is expected for example that an increment in the neutral fraction of the universe leads to a reduction of the observed number of sources, thus supressing the LFs towards higher redshifts (e.g. \citealt{haiman99}). Following this idea, \cite{ouchi10} used the drop in LAEs LF at $\zz=5.7-6.6$ to constrain $x_{\rm HI} \lesssim 0.2 \pm 0.2$  at $\zz=6.6$. A less stringent limit of $x_{\rm HI} \lesssim 0.5$ has been found at the same redshift by \cite{ouchi10} using measurements of LAE clustering. These are marginally consistent with the estimate by \cite{taniguchi05} that at least 20-50 per cent of the volume of the universe needs to be ionized to account for the observed number of LAEs at $\zz=6.5$. The use of number density (e.g. \citealt{malhotra06}) and clustering properties (e.g. \citealt{furlanetto06,mcquinn07}) of LAEs have also been extensively discussed in the literature as tools to probe the EoR.

A large number of authors have modelled LAEs at different redshifts using analytic \citep[e.g.][]{haiman02,kobayashi06,dijkstra07b,kobayashi07,samui09,tilvi09} and semi-numeric \citep[e.g.][]{ledelliou05,ledelliou06, furlanetto06, tasitsiomi06, mcquinn07,iliev08,dayal09,zheng10,dijkstra11,yajima11b} methods. It has been shown that properties like dust content \citep[e.g.][]{kobayashi07,dayal09} and gas inflows \citep[e.g.][]{santos04,dijkstra07a} reduce the observability of LAEs, whereas top-heavy initial mass function \citep[IMF; e.g.][]{malhotra02,dijkstra07c}, clustering of ionizing sources, 'patchy' reionization \citep[e.g.][]{wyithe05,furlanetto06,mcquinn07,iliev08,mesinger08a,mesinger08b,dijkstra11} and gas outflows \citep[e.g.][]{santos04,dijkstra10} improve their observability at each redshift. Most works simulate the scattering of Ly$\alpha$ photons away from the line of sight by suppressing the intrinsic spectrum of an LAE by the optical depth calculated along the line of sight (see equations~\ref{eqn:tau1} and~\ref{eqn:tau2}). Only a few studies perform a full 3D RT (RT) of Ly$\alpha$ photons \citep[e.g.][]{cantalupo05,tasitsiomi06,laursen07,laursen09,kollmeier10,zheng10,barnes11,yajima11b}. 

In this work, we look at the structure in the IGM at $\zz=7.7$, where the current constraints point towards non-negligible neutral fraction in the volume averaged IGM, specifically focusing on the effects of the gas close to the object. This region is important because the inhomogeneities in the gas density play a crucial role in determining the size of the ionized region, especially at very low ionization fractions. The residual neutral hydrogen close to the source, together with the velocity field in this same region strongly affects the scattering pattern of the Ly$\alpha$ photons and their escape towards an observer. Here we investigate these issues for the first time in detail for a large (i.e. $> 100$) sample of objects. LAEs have been detected up to $\zz\sim 7$ \citep[e.g.][]{iye06,ono11}. Because observations in the near future are aiming for $\zz\sim7.7$ \citep[e.g.][]{tilvi10,clement11}, in this study we have chosen to concentrate on objects at $\zz=7.7$.

\cite{zheng10} previously simulated a large sample of LAEs with Ly$\alpha$ RT at moderate spatial resolution in a highly ionized IGM at $\zz=5.7$. Similarly, \citet{laursen11} studied scattering in the post-reionization IGM using hydrodynamical simulations that had much higher spatial resolution for RT. Their work is complementary to our study.

\begin{table*}
\begin{center}
\begin{tabular}{|c|c|c|c|c|c|}
\hline
Model  & $L$ ($\Mpch$) &  number of particles & $\pdm$  ($\msun$) &  $\pgas$ ($\msun$) & $\eta$ ($\kpch$) \\
\hline
L05 & 5 & 2 $\times$ 320$^{3}$ & 3.93 $\times$ 10$^{5}$ & 6.04 $\times$ 10$^{4}$ & 0.78 \\
L10 & 10 & 2 $\times$ 320$^{3}$ & 3.14 $\times$ 10$^{6}$ & 4.83 $\times$ 10$^{5}$ & 1.56 \\
L20 & 20 & 2 $\times$ 320$^{3}$ & 2.52 $\times$ 10$^{7}$ & 3.87 $\times$ 10$^{6}$ & 3.13 \\
L30 & 30 & 2 $\times$ 320$^{3}$ & 8.49 $\times$ 10$^{7}$ & 1.30 $\times$ 10$^{7}$ & 4.69 \\
\hline
\end{tabular}
\caption[Hydro-simulations for LAE modelling.]{Simulation properties. From let to right: model name; comoving box size, L; total number of particles (DM and gas); mass of DM particles, $m_{\rm DM}$; mass of gas particles, 
$m_{\rm gas}$; softening length, $\eta$.}\label{tab:sims}
\end{center}
\end{table*}

In this work we simulate LAEs using galaxies from hydrodynamical cosmological simulations and RT using CRASH$\alpha$ \citep{pierleoni09}. We make surface brightness (SB) maps of the objects and study the effect of IGM structure on the SB profiles. In the following Section, we describe the cosmological simulations and Section~\ref{sec:crasha} briefly introduces the RT code CRASH$\alpha$. Section~\ref{sec:method} explains the pipeline used to simulate a statistically significant sample of LAEs. The results are described in Section~\ref{sec:results} and the parameter study to understand the uncertainties in our work is shown in Section~\ref{sec:paramstudy}. In Section~\ref{sec:reionhistory}, we discuss the effect of IGM structure on the methods used in estimating the mean ionization fraction of the universe. Finally the conclusion are drawn in Section~\ref{sec:discussion}.

\section{Simulations of Galaxy Formation}
\label{sec:sim}

To study high redshift galaxies and their surrounding IGM, we need simulations with a large box size to provide a statistical sample of halos in a wide mass range in different environments. At the same time, high resolution to resolve the halo and the structure in the surrounding IGM is required. A compromise was achieved by using medium range box sizes of 5-30 $\Mpch$ comoving. The simulations used in this paper are described in \cite{maio10}, although additional ones have been run for different box sizes. 

\begin{table*}
\begin{center}
\begin{tabular}{|c|c|c|c|c|}
\hline
 Model & $\Mdm$ ($\times~10^{10}~\msun$)  & $\Mstar$ ($\times~10^{7}~\msun$) &  $\Mgas$ ($\times~10^{8}~\msun$) & $\Zgas$ ($\times~10^{-2}~\Zsun$) \\
\hline
L05 & 0.12 - 2.58 & 6.61 - 256 & 0.03 - 3.63 & 1.20 - 6.87 \\
L10 & 1.33 - 9.70 & 9.06 - 108 & 15 - 110 & 2.02 - 4.80 \\
L20 & 1.84 - 8.57 & 4.59 - 64 & 26 - 110 & 0.76 - 3.52 \\
L30 & 3.58 - 16.6 & 3.54 - 134 & 47.2 - 213 & 0.34 - 2.85 \\
\hline
\end{tabular}
\caption[Halo properties of simulated LAEs at $\zz$=7.7.]{Properties of haloes at $\zz$=7.7. From left to right: name of the model; range of DM halo mass, $\Mdm$; stellar mass, $\Mstar$; gas mass, $\Mgas$; gas metallicity, $\Zgas$.}\label{tab:two}
\end{center}
\end{table*}

The simulations were run using the TREE-PM SPH code Gadget-2 \citep{springel05} modified to include primordial Hydrogen, Helium and Deuterium based chemistry [e$^{-}$, H, H$^{+}$, He, He$^{+}$, He$^{+ +}$, H$_{2}$, H$_{2}^{+}$, H$^{-}$, HeH$^{+}$, D, D$^{+}$, HD] \citep{yoshida03,maio07}, stellar evolution and metal pollution \citep{tornatore07} and fine structure metal transition cooling (O, C$^{+}$, Si$^{+}$, Fe$^{+}$) at $T\;<$ 10$^{4}\;\Kelv$  \citep{maio07,maio09}. Metals produced by AGB stars and supernovae (SNII, SNIa) are spread by supernovae/wind feedback. The IMF is chosen according to the metallicity of the stellar particles, $\Zgas$. We assume that a transition from metal-free/very metal-poor Population III stars to metal-enriched, more standard Population II/I stars takes place whenever the gas reaches a critical metallicity $\Zcrit$, which is determined by the cooling and fragmentation properties of the gas. $\Zcrit$ quoted by different authors range between 10$^{-6}$ to 10$^{-3.5}\;\Zsun$ \citep[e.g.][]{schneider03, bromm03, schneider06,santoro06}. In these simulations we choose $\Zcrit$ = 10$^{-4}\;\Zsun$. Below $\Zcrit$ we assume a Salpeter IMF in the mass range [100,500] $\msun$, while for metallicities above $\Zcrit$ a Salpeter IMF in the mass range [0.1,100] $\msun$ is used.  

The simulations analyzed in this paper have box sizes of $L$ = 5, 10, 20 and 30~$\Mpch$ comoving with 320$^{3}$ particles each in dark matter and gas. The mass of dark matter particles is 3.93 $\times$ 10$^{5}~\msun$ ($L/5~\Mpch$)$^{3}$ and the gas particle mass is 6.04 $\times$ 10$^{4}~\msun$ ($L/5~\Mpch$)$^{3}$. The comoving softening length is 0.78 $\kpch$ ($L/5~\Mpch$)$^{3}$ which is $\sim 1/20$ of the mean inter-particle distance. The details of the simulations are given in Table~\ref{tab:sims}. 

\section{The Radiative Transfer Code CRASH$\alpha$}
\label{sec:crasha}

CRASH$\alpha$ \citep{pierleoni09} is the first RT code for cosmological application where the propagation of Ly$\alpha$ and ionizing photons are coupled (refer to \citealt{yajima11b} for a similar code). The ionizing part is based on CRASH \citep{ciardi01,maselli03,maselli09} which is a 3D ray-tracing grid-based RT code using Monte-Carlo (MC) techniques to sample the probability distribution of several parameters like spectrum of sources, emission direction and optical depth. The ionizing radiation is propagated through an arbitrary static H/He gas density field. Both radiation from point sources and diffuse radiation from ultraviolet background or recombination of H/He gas can contribute to the ionizing flux. 

The total ionizing photon count, $E_{\rm s}$ [phot], from each 
source\footnote{Each galaxy within the hydrodynamical simulations which contains
at least a star forming particle is represented within the RT simulation as a point
source.} $s$ with photon rate $\dot{N}_{\rm ion,s}$ [phot s$^{-1}$] over the whole time $t_{\rm sim}$ of the RT simulation, is distributed among $N_{\rm p}$ photon packets containing ionizing photons of different frequencies depending on the source/background spectrum. The emission of photon packets from each source happens at equally spaced time intervals $dt = t_{\rm sim}/N_{\rm p}$. The direction of emission is assigned by the MC sampling of the angular characteristic of the source. As the packets are propagated through each cell $i$, photons are absorbed according to the cell optical depth $\tau _{\rm c}^{\rm i}$ calculated combining the contributions from HI, HeI and HeII. The probability that a photon in a packet is absorbed is calculated using $P(\tau _{\rm c}^{\rm i})$ = 1 - e$^{\tau _{\rm c}^{\rm i}}$. This changes the temperature and ionization conditions of the cell and the photon distribution in a packet. The trajectory of a packet is followed till all the photons are absorbed or in case of non-periodic boundary conditions, till it reaches the edge of the box. 

CRASH is modified to follow the time evolution of Ly$\alpha$ photons through an evolving ionization configuration of gas. A statistical approach to the Ly$\alpha$ RT is adopted using pre-compiled tables from MCLy$\alpha$ \citep{verhamme06}. Ly$\alpha$ radiation is emitted by both point sources and recombination of ionized gas. Just like in the ionizing case, the Ly$\alpha$ photons produced by each point source $s$ with photon rate $\dot{N}_{\rm Ly\alpha,s}$ over the entire simulation time $t_{\rm sim}$ is divided between Ly$\alpha$ photon packets.  $N_{\rm p,Ly\alpha }$ Ly$\alpha$ photon packets are emitted by each source $s$ at regular time intervals $dt_{\rm Ly\alpha} = t_{\rm sim}/N_{\rm em,l}$ in random directions. $N_{\rm em,l}$ is the number of times the Ly$\alpha$ photons are propagated during the RT simulation. When a packet enters a cell, depending on the conditions in the cell, the photons are either absorbed or scattered. In case of scattering, a new wavelength and the time of escape from this cell is obtained from the pre-calculated tables. The typical time of escape from a cell of size 1 kpc, temperature 10 K and number density 0.01 cm$^{-3}$ giving an optical depth of $\sim 6 \times 10^{7}$  is about $3 \times 10^{5}$ years. \citep[e.g.][]{adams75,dijkstra08} Photons from the recombination in the gas are added to the packets. 

Dust is an important factor which determines Ly$\alpha$ radiation transport \citep[e.g.][]{neufeld91}. Dust optical depth is calculated as a fraction $f_{\tau}$ of the Ly$\alpha$ optical depth. $f_{\tau}$ is determined as $f_{\tau}= m_{\rm p/dust} \times f_{\rm H/dust} \times \sigma_{\rm dust} \times \rho_{cell} \times d_{\rm cell}$ where proton-to-dust mass ratio $m_{\rm p/dust}$ = $5.0 \times 10^{-8}$, gas-to-dust ratio $f_{\rm H/dust}$ = $5 \times 10^{-3}$ \citep{verhamme06}, dust absorption cross section $\sigma_{\rm dust}$ = $2  \times \pi \times r_{\rm dust}^{2}$ for a dust grain of radius $r_{\rm dust}$ = $2.0 \times 10^{-6}$cm. $\rho_{cell}$ is the gas density in the cell and $d_{\rm cell}$ is the distance traveled within the cell by the photon. In high density regions, where Ly$\alpha$ photons have a high probability of scattering multiple times within the cell volume, the probability of dust absorption gets high, with the consequence that the effects due to dust scattering on Ly$\alpha$ RT are negligible. For example, assuming that the probability of scattering of Ly$\alpha$ photons by dust is equal to that of absorption, in 20 photon-dust interactions, only 1 in $10^{6}$ photons would continue to scatter without being absorbed. Thus in high density regions, we neglect dust scattering.

For more details, we refer the reader to the original papers. 

\section{Simulating Ly$\alpha$ Emitters}
\label{sec:method}

Our plan is to study how the observability of LAEs is affected by transmission through the IGM. For this, we calculate Ly$\alpha$ SB profiles of a large number of simulated LAEs at $\zz=7.7$ covering a wide range of dark matter halo masses, thus sampling an equally wide range in IGM environments. The redshift was chosen as observational efforts are underway to investigate LAEs at $\zz=7.7$ \citep[e.g.][]{tilvi10,clement11}. As at this relatively high $z$ the IGM is expected to be still substantially neutral, a full 3D RT approach is necessary to investigate the observability of LAEs.

The method adopted to simulate LAEs is to extract a cube around each dark matter halo from the snapshot of the simulation of galaxy formation, grid the density and velocity fields, get the spectrum produced by its stellar population from STARBURST99 \citep{leitherer99} using the halo properties, input all the details to CRASH$\alpha$ along with the default values (refer to Section~\ref{sec:rtsim}) for temperature and ionization fields and run the RT simulation to obtain Ly$\alpha$ photons escaping the simulation cube. We can use the details of the photons exiting each of the six faces of the cube to make SB maps. Each step is explained in detail in the following Sections.

\subsection{Extracting Halos and Gridding}
\label{sec:grid}

\begin{figure}
\centering
\includegraphics[width=85 mm,height=170mm]{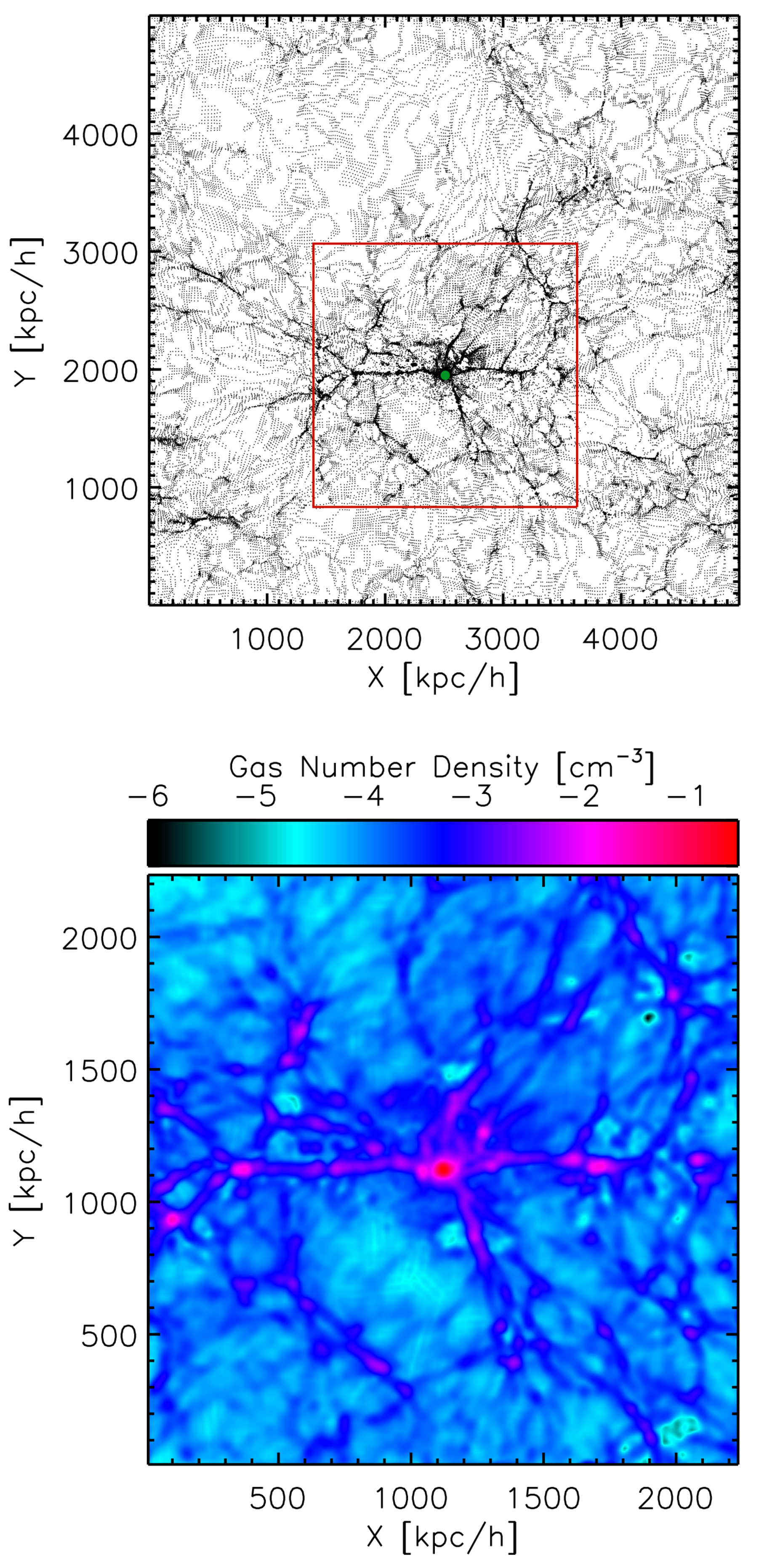}
\caption[Extraction and gridding of a cube around a halo.]{Example of the extraction of a cube around a DM halo and the gridding of the gas density and velocity fields for CRASH$\alpha$. Top: The gas distribution within a thin slice of the simulation box L05 centred on the CoM (shown as a green dot) of the most massive halo at $\zz=7.7$ in comoving units of length. The comoving thickness of the slice is 10 $\kpch$. The cube to be extracted around the object for the RT simulation is shown with a solid red square. Bottom: The gridded gas number density field (in logarithmic scale) corresponding to the cube in the top panel.}
\label{fig:slicecube}
\end{figure}

A snapshot of the hydrodynamical simulation is taken and the properties of the dark matter haloes are obtained using the Friends-of-Friends (FoF) routine \citep{davis85}. The centre-of-mass (CoM) of the dark matter (DM) halo, the DM, gas and stellar mass, and mean gas metallicity are used in the rest of the analysis. The mean gas metallicity is defined as the fraction of gas mass in metals. A summary of the DM halo properties for  the four different simulations used in this paper are given in Table \ref{tab:two}. In our study, we choose a subset of the resolved haloes \citep[i.e. with mass $ \Mdm > 100$~$\pdm$; e.g. ][]{trenti10} from L05-30 to get a fair sampling of the DM mass range 10$^{9-11} \msun$. This results in $\sim 130$ haloes.

In our simulations, we resolve the structure in the IGM, but we do not properly resolve the ISM of the individual galaxies. To mimic the absorption/scattering of radiation within the ISM, we use the commonly adopted prescription based on the escape fraction of ionizing photons, $\fescionism$, and the escape fraction of Ly$\alpha$ photons, $\fesclyaism$ (see Sec.~\ref{sec:escfrac}). Thus we need to exclude from our RT calculations those cells which represent the ISM, to avoid accounting for their effect twice. We obtain this by removing all cells with a density larger than $0.03$ cm$^{-3}$. Even values up to 0.05~cm$^{-3}$ are acceptable, but might show slight dependence on the gridded density resolution especially for high resolution gridding in the case of small haloes (see discussion in the next paragraph). For the case of $0.03$ cm$^{-3}$, the effect of density resolution is negligible in the range of simulations we deal with in this work. Values $< 0.03$ cm$^{-3}$ give a similar ionization structure but lead to loss of some of the IGM gas. As a precaution to avoid removal of gas from high density regions not associated with the specific dark matter halo, we restrict the removal of cells to a distance of $\sim 0.7 \times r_{\rm 200}$ from the source. The precise choice of this radial distance does not 
affect our main results, as long as it is less than the nearest massive dark matter halo in the cube.

A cube around the CoM of the halo is extracted with a side of 35 $\times$ $r_{\rm 200}$, where $r_{\rm 200}$ is the radius in comoving units at which the mean DM density inside the sphere is 200 times the critical density of the universe at that redshift. This is the smallest box containing the HII region produced by the galaxy at its equilibrium, so that we get the highest possible spatial resolution for the gridding of density and velocity fields for our RT simulations. The exact value of the cube dimension though does not affect our main results. A cube size in the range $25-50~\times~r_{\rm 200}$ gives similar results, but the lower bound is too close to the edges of the ionized region, while larger boxes would have lower spatial resolution at fixed grid size, which is undesirable. 

Once the cube is extracted, the gas density and velocity fields are gridded as input for CRASH$\alpha$. For gridding, we use GadgettoGrid \citep{pakmor10}. The code distributes the particle mass and peculiar velocities using an SPH kernel to 64 neighboring particles and then grids the fields. The default grid size used is 256$^3$, which is set by memory and runtime constraints of CRASH$\alpha$, although different grid sizes have been used for testing purposes. An important caveat is that, because the grid resolution is a factor of the virial radius of the halo, this results in a lower physical resolution for higher mass objects. 

Figure~\ref{fig:slicecube} shows a slice through the simulation box. The top panel refers to the gas number density in the box with the to-be-extracted cube around a halo marked by a solid red line and the CoM of the halo marked by a green dot. The extracted density field is gridded and shown in the bottom panel. Note that the gas density in the IGM is not uniform and the gridded density field has a range in values spanning many orders of magnitude. 

\subsection{Luminosity of Stellar Sources}
\label{sec:luminosity}

\begin{figure*}
\centering
\includegraphics[width=120mm,height=65mm]{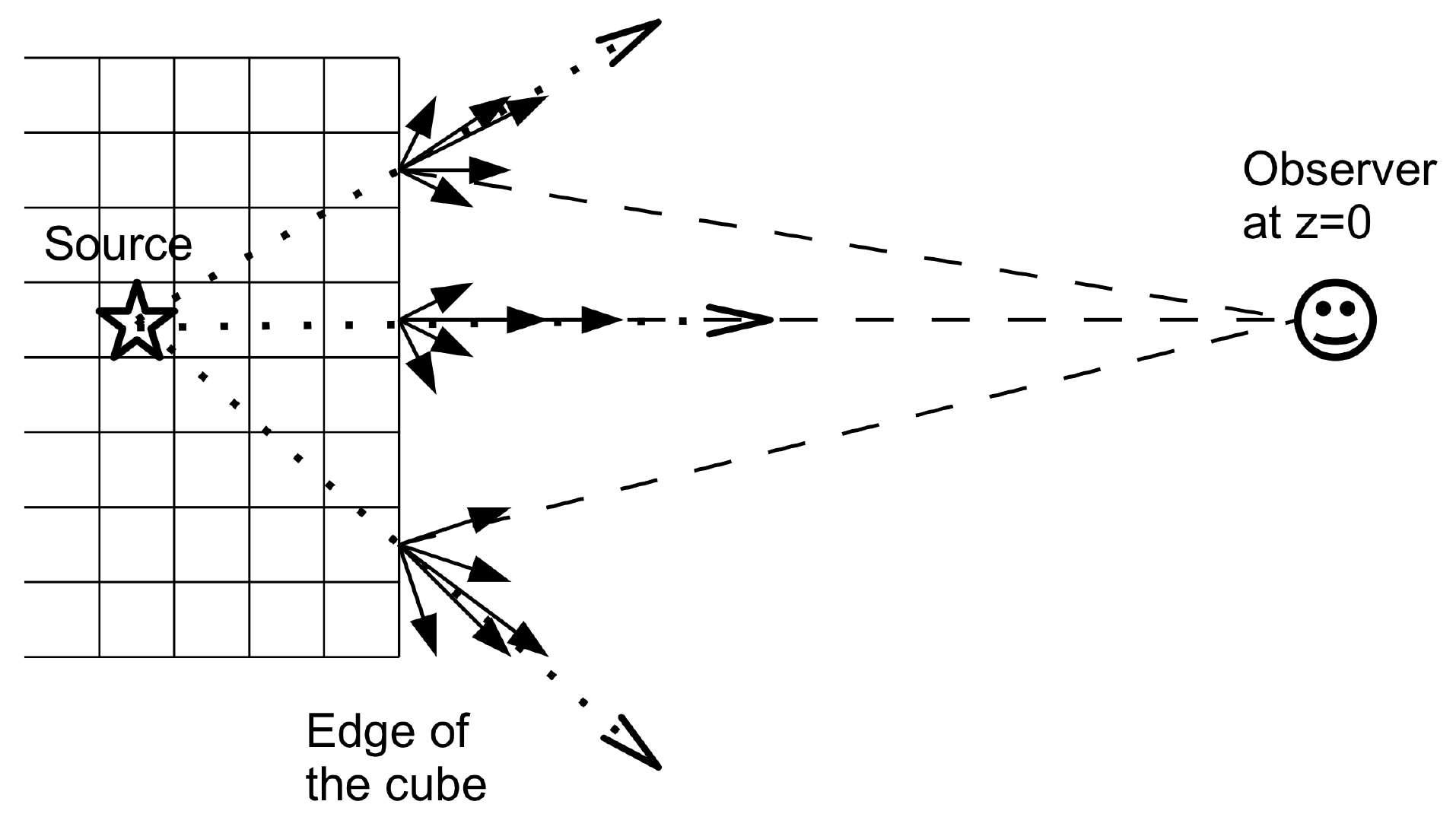}
\caption[Sketch of the method used to calculate the SB.]{Sketch of the method used to calculate the SB. The star represents the pixel with the stellar source. The dotted lines represent the radial direction connecting each pixel on the edge of the box to the source. The solid lines with arrows represent the directions of photon packets within such pixels. Dashed lines connect the pixels to the observer at $z=0$.}
\label{fig:sbsketch}
\end{figure*}

As mentioned in the previous Section, we obtain the halo properties using the FoF and use the associated IMF, stellar mass and metallicity to calculate the corresponding ionizing and Ly$\alpha$ photon rate. To this purpose, we use STARBURST99 \citep{leitherer99}, which is a population synthesis code that provides the spectrum of a stellar population given an IMF, stellar mass, metallicity and age. 

The lowest metallicity available in STARBURST99 is $\Zgas=0.0004 \sim 0.02~\Zsun$ for Padova tracks. We choose the Padova original stellar evolution tracks instead of Padova AGB tracks as the Hubble age at $\zz=7.7$ is less than $1$ Gyr, allowing us to ignore the contribution of AGB stars to the spectrum. The average ionizing flux for the same stellar mass is mildly higher for lower metallicities, i.e.~$\sim 17\%$ higher for $\Zgas=3 \times 10^{-3} ~\Zsun$ (refer to \citealt{schaerer03} for low metallicity values). Since our metallicities are above $\Zcrit$, we use a Salpeter IMF in the mass range [0.1,100] $\msun$ in the STARBURST99 models, consistently with our cosmological simulations. The spectrum is scalable with respect to the stellar mass of the halo. 
Therefore, we choose $\Mstar=10^{6}\;\msun$ as the default stellar mass for the STARBURST99 model and normalize the photon count to this stellar mass. We also choose instantaneous star formation mode for computing the spectrum. The mean ionizing photon rate $\dot{N}_{\rm ion}$ was obtained by averaging the number of photons emitted over $t = 2 \times 10^{8}$ years in the wavelength range $[91,912]\rm \AA$. The time $t$ was chosen as the cumulative number of ionizing photons reaches convergency. Thus, we obtain a normalized ionizing photon rate $\dot{N}_{\rm ion}=7.77 \times\;10^{50}~(\Mstar/10^{6}\;\msun)$ phot~s$^{-1}$. Only a fraction of these ionizing photons reaches the IGM, the rest is converted to Ly$\alpha$ photons in the ISM. Therefore, the ionizing photon rate reaching the IGM, $\dot{N}_{\rm ion}^{\rm esc}$, is:
\begin{equation}
\dot{N}_{\rm ion}^{\rm esc} = \fescionism \times \dot{N}_{\rm ion},
\end{equation}
where $\fescionism$ is the escape fraction of ionizing photons from the ISM. Our reference value is $\fescionism=0.02$ \citep{gnedin08}. In Section~\ref{sec:escfrac} we discuss more extensively this choice.

The calculation of the Ly$\alpha$ photon rate $\dot{N}_{\rm Ly\alpha}$ is more complicated. There are three different components considered in our work - stellar continuum  $\dot{N}_{\rm Ly\alpha,\star}$, nebular continuum $\dot{N}_{\rm Ly\alpha,\;neb}$ and recombination in the ISM $\dot{N}_{\rm Ly\alpha,\;ISM}$:
\begin{equation}
\dot{N}_{\rm Ly\alpha} = \dot{N}_{\rm Ly\alpha,\star}+\dot{N}_{\rm Ly\alpha,\;neb}+\dot{N}_{\rm Ly\alpha,\;ISM}.
\end{equation}

$\dot{N}_{\rm Ly\alpha,\star}$ and $\dot{N}_{\rm Ly\alpha,\;neb}$ are calculated from the spectrum by averaging over $t$ at $1215.67 \rm \AA$ with a line width of 2 $\rm \AA$, which is the typical intrinsic line width of LAEs \citep{partridge67}. In STARBURST99 spectra, the nebular continuum is calculated by converting all the ionizing photons into nebular lines except for Ly$\alpha$. Therefore, $\dot{N}_{\rm Ly\alpha,\;neb}$ gives the upper limit of the contribution from other nebular lines to this wavelength range.

$\dot{N}_{\rm Ly\alpha,\;ISM}$ is instead computed from the ionizing photon rate as \citep[also see e.g.][]{schaerer03}: 
\begin{equation}
\dot{N}_{\rm Ly\alpha,\;ISM}=0.68 \times \dot{N}_{\rm ion}\times(1-\fescionism).
\label{eqn:lyaism}
\end{equation}
Assuming our reference value of $\fescionism=0.02$, the rate of Ly$\alpha$ photons are $\dot{N}_{\rm Ly\alpha,\star} = 6.94 \times\;10^{48} ~(\Mstar/10^{6}\;\msun)$ phot~s$^{-1}$,  $\dot{N}_{\rm Ly\alpha,\;neb} = 2.83 \times\;10^{47} ~(\Mstar/10^{6}\;\msun)$ phot~s$^{-1}$ and $\dot{N}_{\rm Ly\alpha,\;ISM} = 5.07 \times\;10^{50} ~(\Mstar/10^{6}\;\msun)$ phot~s$^{-1}$ (for an ionized gas temperature of $10^4$ K). Thus, $\dot{N}_{\rm Ly\alpha}=5.15 \times\;10^{50} ~(\Mstar/10^{6}\;\msun)$ phot~s$^{-1}$ and our total Ly$\alpha$ photon rate is completely dominated by photons that were emitted as recombination radiation in the ISM.

Finally, the $N_{\rm Ly\alpha}^{\rm esc}$ which escapes the galaxy into the IGM after absorption by dust can be defined as:
\begin{equation}
\dot{N}_{\rm Ly\alpha}^{\rm esc} = \fesclyaism \times \dot{N}_{\rm Ly\alpha},
\label{eqn:dust}
\end{equation}
where $\fesclyaism$ is the escape fraction of Ly$\alpha$ after absorption by dust in the galaxy. Our fiducial value is $\fesclyaism = 0.3$ \citep{dayal09}. In Section~\ref{sec:escfrac} we discuss more extensively this choice. $\dot{N}_{\rm Ly\alpha}^{\rm esc}$ can also be converted into a luminosity, $L_{\rm Ly\alpha}^{\rm esc}$, by convolving with the spectrum of Ly$\alpha$ photons.

Our value of $\dot{N}_{\rm ion}$ is a factor of two lower than the numbers quoted by e.g. \cite{dayal09} and \cite{zheng10} due to differences in the calculation of the ionizing photon rate. For example, \cite{dayal09} uses the IMF mass range of [1,100] $\msun$ rather than [0.1,100] $\msun$ employed in our simulations. Thus, their values would be comparable to ours for $\fescionism = 0.04$ and $\fesclyaism=0.6$.

\begin{figure}
\centering
\includegraphics[width=85mm,height=90mm]{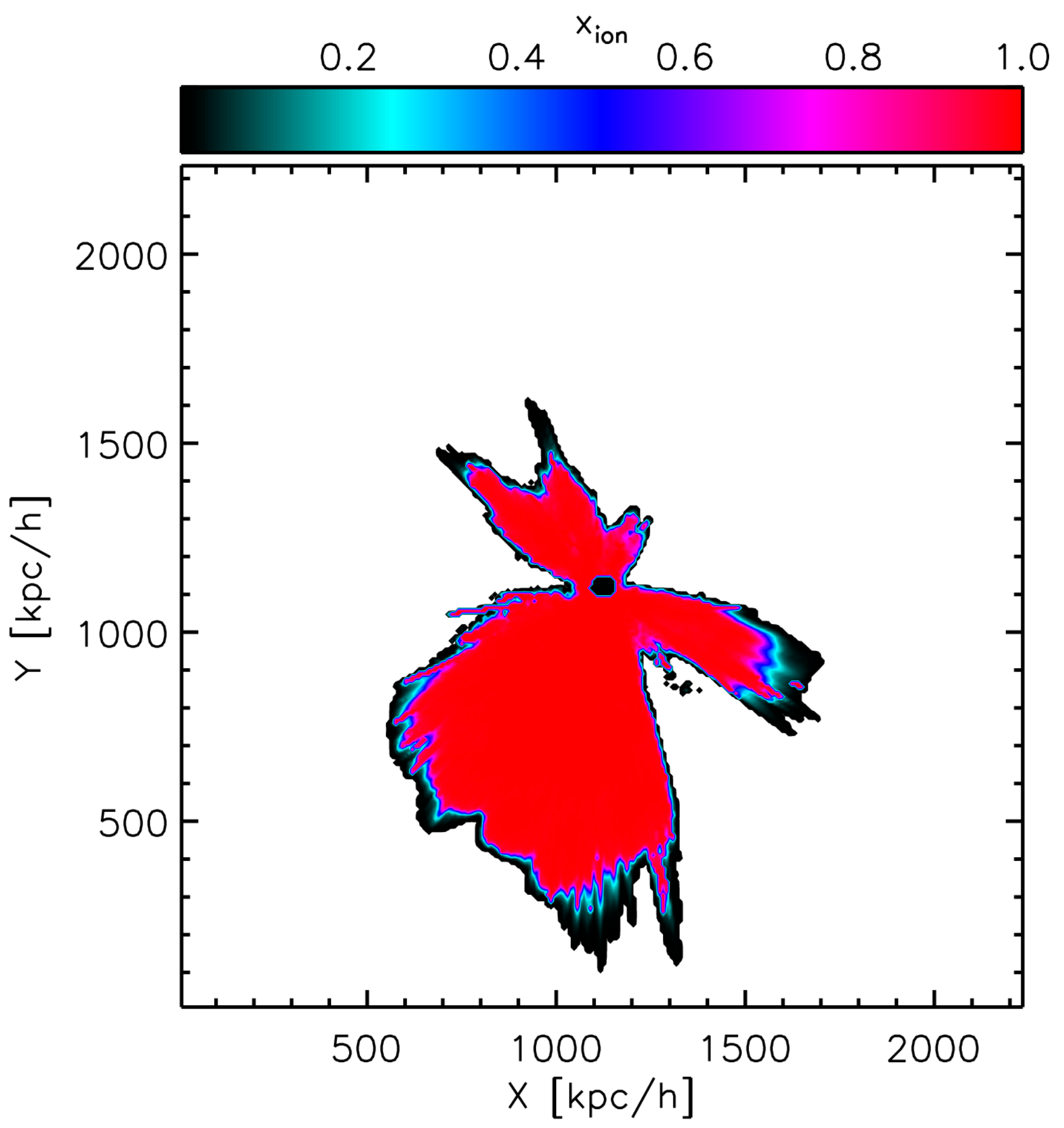}
\caption[The ionization region around the most massive object in L05]{The ionized region around the most massive object in L05 shaped by the density structure of the IGM close to the source. The degree of ionization of the gas is shown by the color bar at the top.}
\label{fig:ionrun}
\end{figure}

\subsection{CRASH$\alpha$ Input and Output}
\label{sec:rtsim}

The inputs for a CRASH$\alpha$ run are the gridded density, velocity and temperature fields (Sec.~\ref{sec:grid}) and a source distribution with photon rate and spectrum (Sec.~\ref{sec:luminosity}). In our reference simulations, we adopt uniform initial fields of ionization, $\ionfrac = n_{\rm HII}/n_{\rm H}=0$, and temperature, $T = 100$~K. A discussion on the different choices for these values is given in Section~\ref{sec:ionization}. The duration of the RT simulation is set to $2 \times 10^{8}$ years. The source locations and luminosity were calculated as explained in the previous Section. Other simulation parameters, $N_{\rm p}$ = $10^{6}$, $N_{\rm p,Ly\alpha}$ = $10^{4}$ and $N_{\rm em,l}$ = 500, were determined using resolution tests.

Throughout the simulation, CRASH$\alpha$ collects photon packets exiting the box, recording their frequency, time, position and directional information, which can be used to quantify observed properties of the source such as SB maps and spectra. Each cell on a side of the simulation cube emits photon packets in different directions. Just like the photon counts, the true distribution of angular directions of the photons exiting from the cells are also sampled by the photon packets. Each direction points towards a different observer thus making each cell (3D) act as a pixel (2D) in the plane perpendicular to the direction of the observer. Here we define the direction of the observer as the one perpendicular to the plane of the side of the cube. By this definition we get six different observer directions for each source. Both SB maps and spectra can be calculated for each observer direction. 

The SB $SB_{\rm em}$ of a pixel of size $p_{s}$ due to a source of luminosity $L$ at a distance $d$ (in Euclidian geometry) is given by:
\begin{equation}
SB_{\rm em} = \frac{L}{4 \pi d^{2}} \times \frac{1}{(p_{s}/d)^2} = \frac{L}{4 \pi p_{s}^{2}}.
\end{equation}
 The photons, after travelling through the expanding universe, arrive at the observer at $\zz=0$. In a cosmological context, the observed SB $SB_{\rm obs}$ of a source of luminosity $L$ at redshift $\zz=\zz_{s}$ observed in a pixel of proper length $p_{s}$ is \citep{tolman34}: 
\begin{equation}
SB_{\rm obs} = \frac{L}{4 \pi D_{L}^{2}} \times \frac{1}{(p_{s}/D_{A})^2} = \frac{L}{4 \pi p_{s}^{2} (1+\zz_{s})^{4}} = \frac{SB_{\rm em}}{(1+\zz_{s})^{4}},
\label{eqn:sb}
\end{equation}
where $D_{L}$ is the luminosity distance of the source and $D_{A}$ is its angular diameter distance. Therefore, to calculate the observed SB in a pixel for a specific observer, we need the number of photons in each cell travelling in the direction of the observer. The sampling of the angular distribution of photons depends on the number of photon packets in each cell. But the average number of photon packets in each cell of our simulation is too low ($\sim 12$) for a smooth sampling of the underlying angular directional distribution. Thus we calculate the angular directional distribution using all the photon packets on a side of the box, by assuming that all cells on that side sample the same angular directional distribution with respect to the true north for each cell, which is defined as the radial direction from the source to the cell. 

Figure~\ref{fig:sbsketch} shows the sketch of the method. The star represents the cell with the stellar source. The dotted lines represent the radial direction connecting each cell on the edge of the box to the source (i.e. true north for each cell). The solid lines with arrows represent the directions of photon packets exiting such cells. The dashed lines from each cell to the observer represent the observer direction, which for $\zz_{s} = 7.7$ is perpendicular to the side. The packets which are not scattered by the IGM follow the radial direction from the source (i.e. true north), while the scattered photon packets sample random directions. To calculate the $SB_{\rm obs}$ we need to estimate the number of photons reaching the observer at $z=0$, i.e. traveling along the dashed lines. To calculate the underlying angular direction distribution, we grid the directions of all photon packets (short solid arrows) in each cell, correcting for the different true north direction of each cell (dotted arrow). The gridded distribution of angles with respect to their true north in all the cells on a side are added and normalized using the total number of photon packets exiting the side. This normalized gridded angular distribution is then imposed on the photon count in each cell to obtain the number of photons in each pixel going in the direction of the observer.

\begin{figure*}
\centering
\includegraphics[height=120mm,width=170mm]{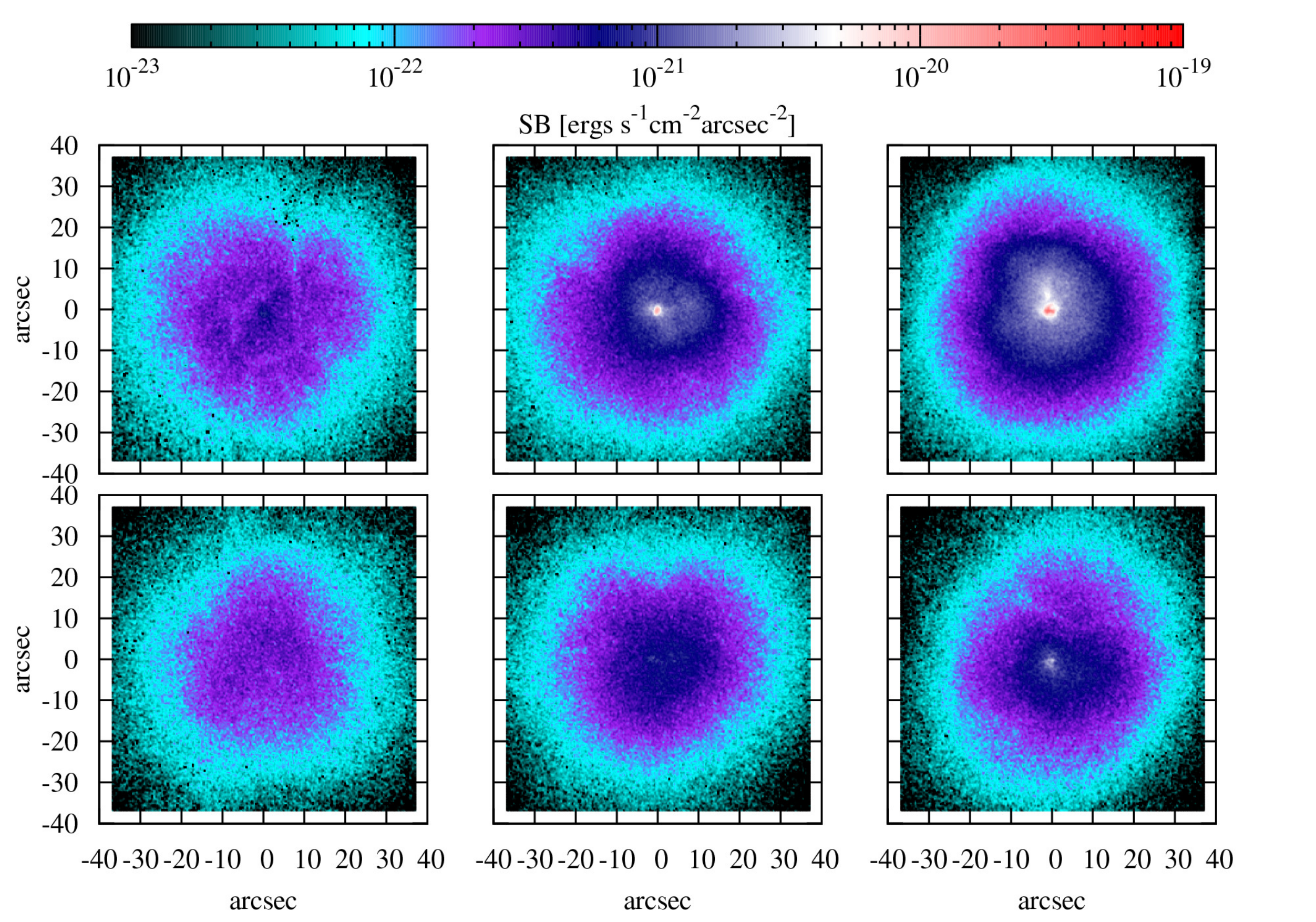}
\caption[Surface brightness maps of the most massive object in L05]{Contour plot showing the six different SB maps for the same object as obtained from the six sides of the simulation cube. The object in this simulation is the most massive object in L05, the same one as in Figure~\ref{fig:slicecube}. The SB values in the pixels are shown in the color bar at the top. See text for details.}
\label{fig:run1sb}
\end{figure*}

The total energy of photons travelling in a given direction within a pixel in each second, $L_{\rm pixel}$, is given by:
\begin{equation}
L_{\rm pixel} = \frac{L}{4 \pi d^{2}} \times p_{s}^{2},
\end{equation}
which is calculated by adding up the energy of all the photons in the pixel in the direction $\theta \le  p_{s}/d$. $d$ is the proper distance of the cell to the source (at the centre of the simulation box). $L_{\rm pixel}$ is converted to the units of SB, i.e. erg~s$^{-1}$~cm$^{-2}$~rad$^{-2}$, using the formula:
\begin{equation}
SB_{\rm em} = L_{\rm pixel} \times \frac{d^{2}}{p_{s}^{4}}.
\end{equation} 
This is then converted to the $SB_{\rm obs}$ dividing by $(1+z)^{4}$ as in equation~(\ref{eqn:sb}). The $SB_{\rm obs}$ is then converted to flux, $F$, as:
\begin{equation}
F = SB_{\rm obs} \times \frac{p_{s}^{2}}{D_{A}^{2}},
\label{eqn:fobs}
\end{equation} 
where $p_{s}/D_{A}$ is the pixel size of the observer in radians and $D_{A}$ is the angular diameter distance of the source. Adding up the flux in all pixels, we can obtain the observed source luminosity $L_{\rm obs}$ as:
\begin{equation}
L_{\rm obs}=4 \pi D_{L}^{2} \times F,
\label{eqn:lobs}
\end{equation} 
where $D_{L}$ is the luminosity distance of the source.

An important technical caveat is that our method for computing the SB differs from the often used 'next-event estimator'  approach which has been adopted in other studies \citep[e.g.][]{tasitsiomi06,laursen09,zheng10,kollmeier10,barnes11}. Idealized test cases though show that the method perfectly reproduces the SB profiles of the Rybicki-Loeb halo \citep{loeb99}, and also those produced in cases where a spherical symmetric ionized bubble is carved out of the neutral expanding IGM (as in Dijkstra \& Wyithe 2010). This approach though is less appropriate for cases where scattering occurs in only a few (neutral) clumps embedded in an otherwise fully transparent medium. However, the method works well in test cases similar to the configurations we are studying in our simulations, and thus we consider the main results of this study to be robust.

\begin{figure}
\centering
\includegraphics[height=68mm,width=85mm]{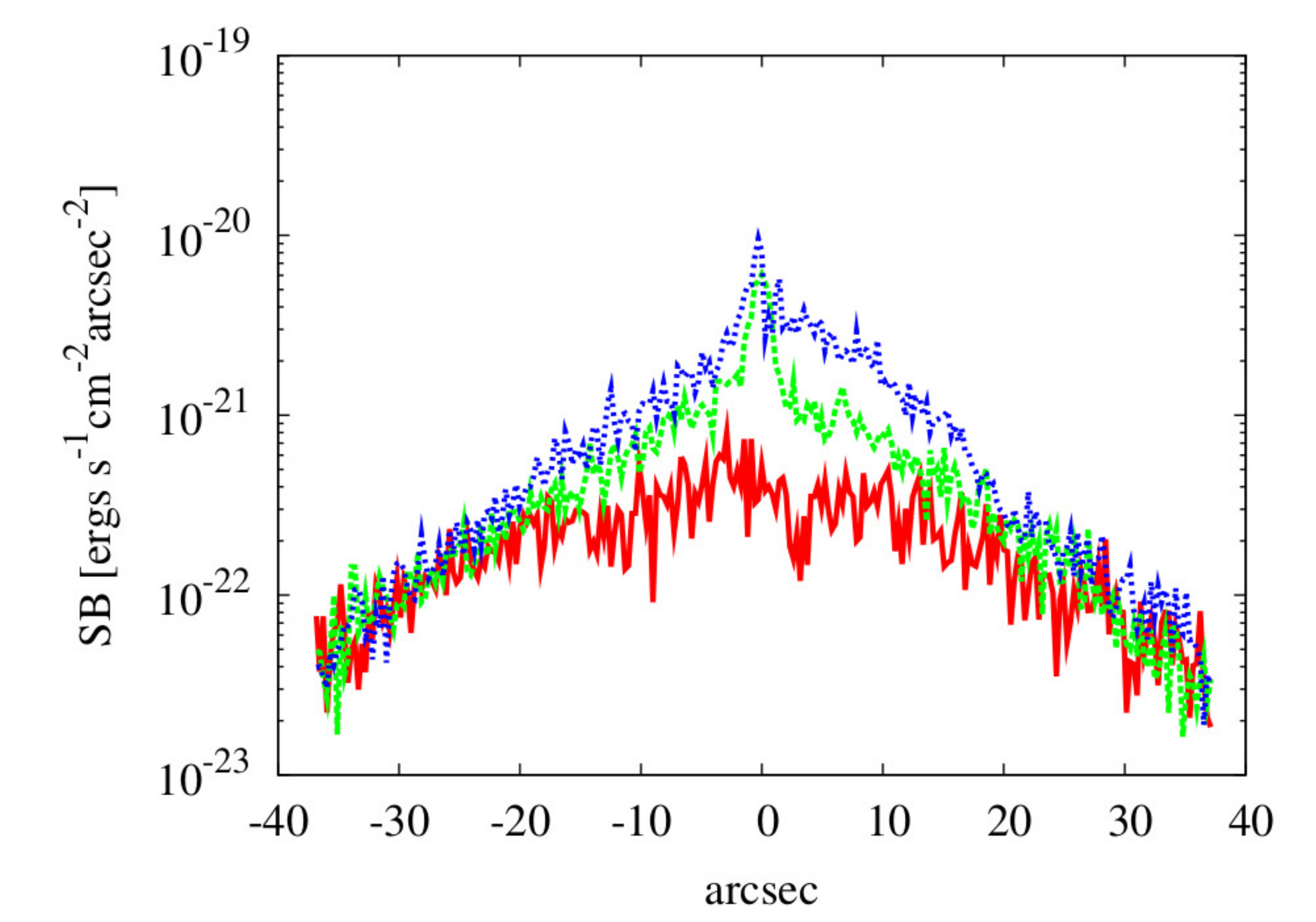}
\caption[Cross-section of the SB map.]{A cross section of the SB profile for the maps in the top row of Figure~\ref{fig:run1sb}. The lines are for SB values along the y axis (i.e. x values are 0) of the maps. The solid red, dashed green and dotted blue lines refer to the top-left, top-middle and top-right maps, respectively.}
\label{fig:run1sbprofile}
\end{figure}

\section{Results}
\label{sec:results}

Using the pipeline outlined in the previous Section, we simulate $\sim 130$ LAEs from L05-30 to evenly sample the DM mass range 10$^{9-11}~\msun$. Table~\ref{tab:two} contains the properties of the haloes used in this work. The simulations were run with our default parameters (see Sec.~\ref{sec:rtsim}). Dependence of the results on the choice of the parameters is discussed in Section~\ref{sec:paramstudy}. Before trying to understand the properties and statistical trends of $130$ galaxies, we focus on understanding how the IGM close to the objects affects the appearance of a single LAE. We investigate the ionization structure in the IGM and how it affects the 
SB maps. Having understood the effects on an individual galaxy, we study the behaviour of the whole sample and the trends shown in observability and Ly$\alpha$ escape fractions due to the IGM.

\subsection{Behaviour of an Individual Galaxy}

To understand how the density field in the IGM around the source affects the shape of the ionized region, which in turn will reflect on the propagation of the Ly$\alpha$ photons, we look at the ionization structure around the most massive dark matter halo in L05 (same galaxy as in Fig.~\ref{fig:slicecube}) of DM mass of $2.6 \times 10^{10}~\msun$ and stellar mass of $3.6 \times 10^{8}~\msun$. Figure~\ref{fig:ionrun} shows the mid-plane of the ionization structure at the end of the RT simulation. We can see that the ionized region is not spherical and its edges are shaped by the high density regions (e.g. filaments) in the IGM. Ly$\alpha$ photons propagating through the ionized gas will travel a large distance unscattered while being redshifted away from the Ly$\alpha$ rest wavelength, thus improving the chances of escape from the neutral IGM beyond the ionized region. Ly$\alpha$ photons encountering a high density filament instead will undergo several scatterings before being able to eventually escape the simulation box. Because of the different paths followed by the photons, we expect different SB distributions and spectra, depending on the viewing direction. This can be clearly seen in Figure~\ref{fig:run1sb}, where the SB maps obtained from the six sides of the simulation box are shown for the same object. 

The differences in the maps reflect the structure in the IGM. In general, the lesser the photons are scattered before reaching the edge of the box, the more concentrate the SB profile is. In this case, the central pixels also have a much higher value of SB compared to cases in which the photons are scattered by high density neutral gas. Note that, as a result of the very inhomogeneous structure of the IGM, the maximum SB value in the image differs by  more than an order of magnitude for different lines-of-sight. This is more evident in Figure~\ref{fig:run1sbprofile}, which shows a cross section of the SB maps for the three directions shown in the top row of Figure~\ref{fig:run1sb}. Plotted are the SB values in the map for an x axis value of 0 against the y axis of the map. As noted earlier, the difference in the value of the central pixels in the  maps is more than an order of magnitude for different viewing directions of the same object.

\begin{figure}
\centering
\includegraphics[width=80mm,height=60mm]{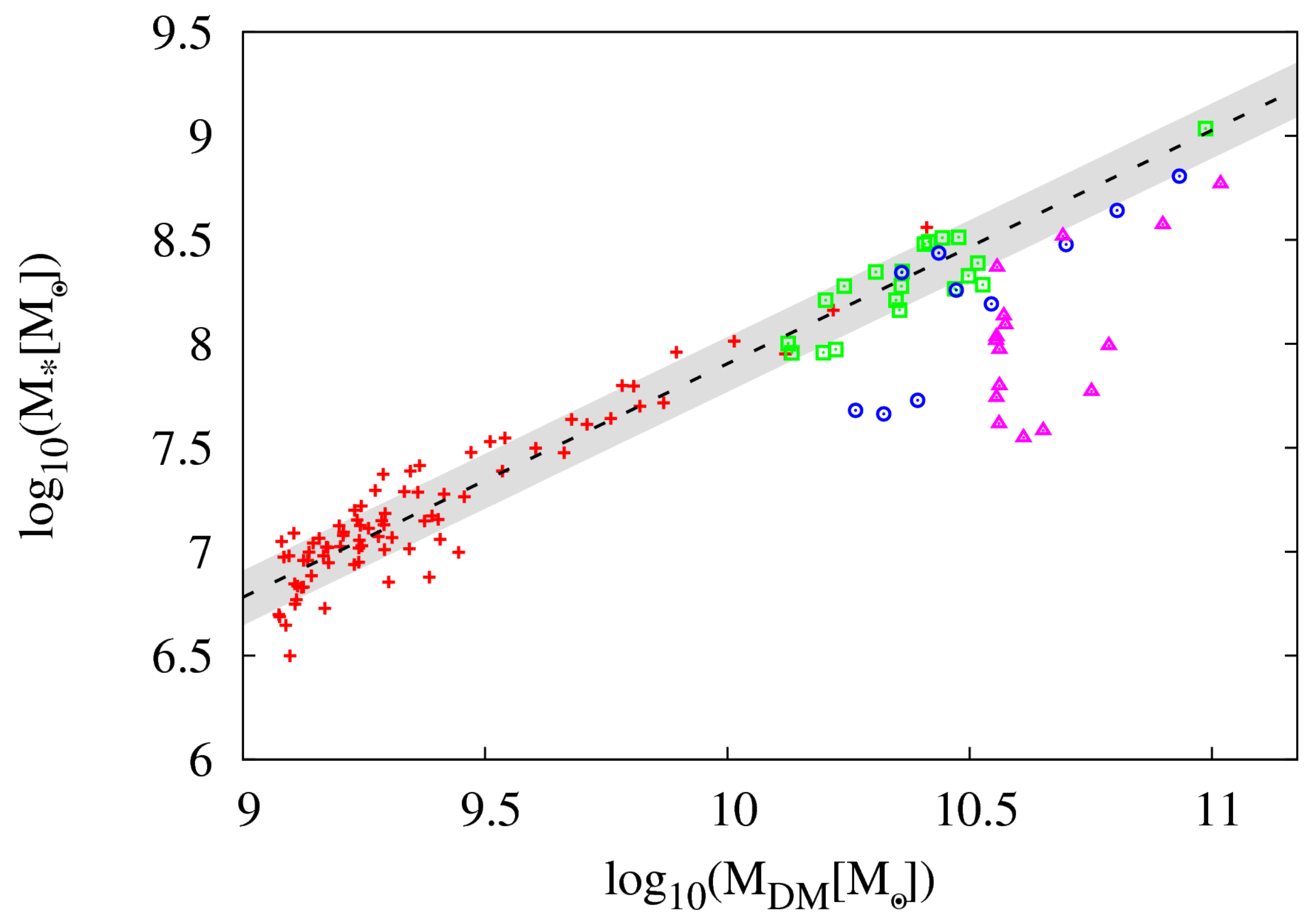}
\caption[Stellar mass -dark matter mass correlation.]{Stellar mass of the objects plotted against their DM mass. The objects from different simulations are marked as - L05 (red crosses), L10 (green squares), L20 (blue circles) and L30 (pink triangles). The best fit line (dashed) is obtained using only haloes from L05 and L10, and it has a log-log slope of 1.12 with a standard deviation $\sigma$=0.13 dex. The $\sigma$ region is shaded in light grey.}
\label{fig:stardm}
\end{figure}

\begin{figure}
\centering
\includegraphics[width=85mm,height=60mm]{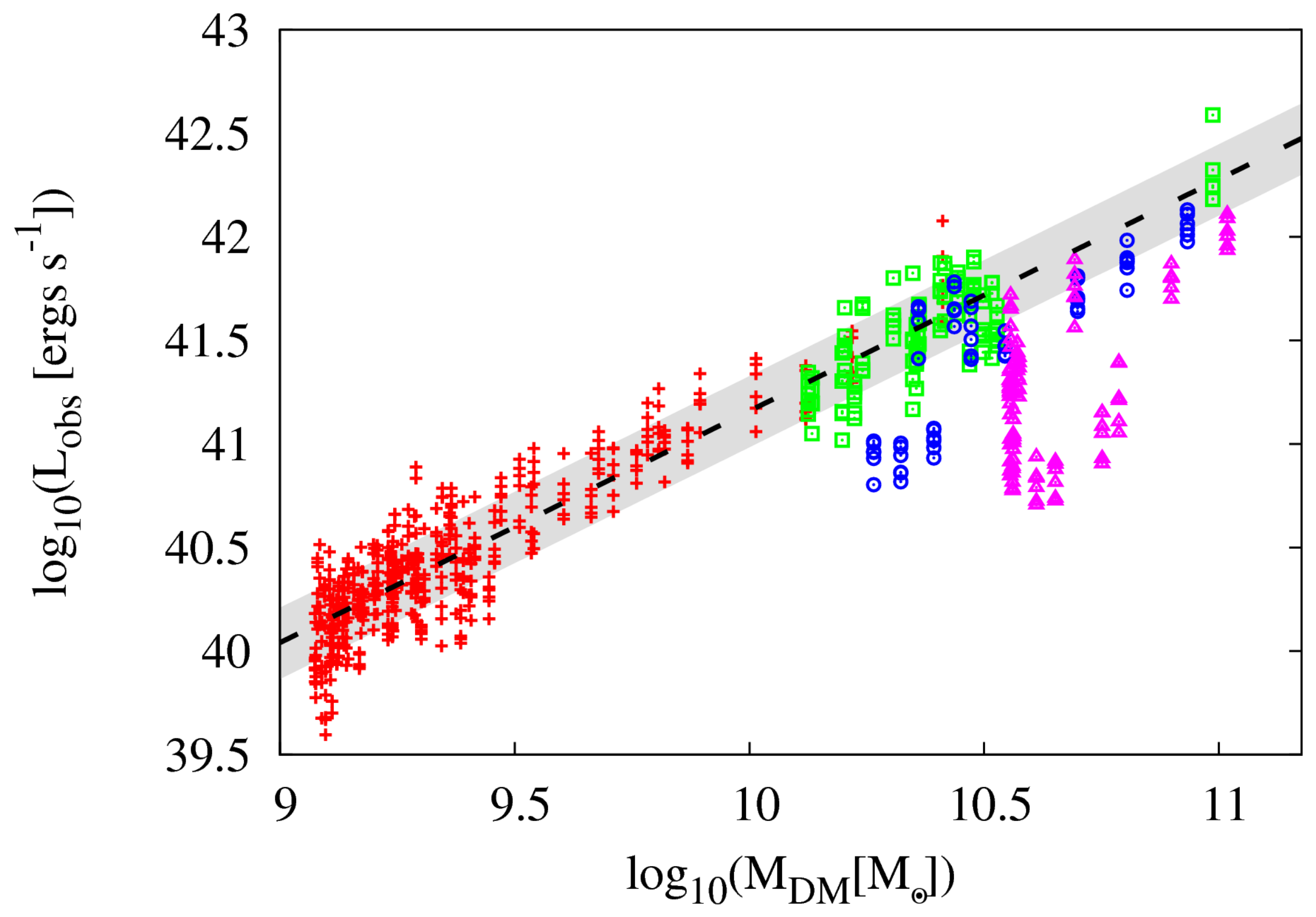}
\caption[Observed luminosity - DM mass correlation.]{Observed Ly$\alpha$ luminosity of the objects plotted against their DM mass. The objects from different simulations are marked as - L05 (red crosses), L10 (green squares), L20 (blue circles) and L30 (pink triangles). The best fit line (dashed) is obtained using only haloes from L05 and L10, and it has a log-log slope of $\sim 1.12$ with a standard deviation $\sigma$=0.17 dex. The $\sigma$ region is shaded in light grey.}
\label{fig:lumdm}
\end{figure}

\begin{figure}
\centering
\includegraphics[width=90mm,height=70mm]{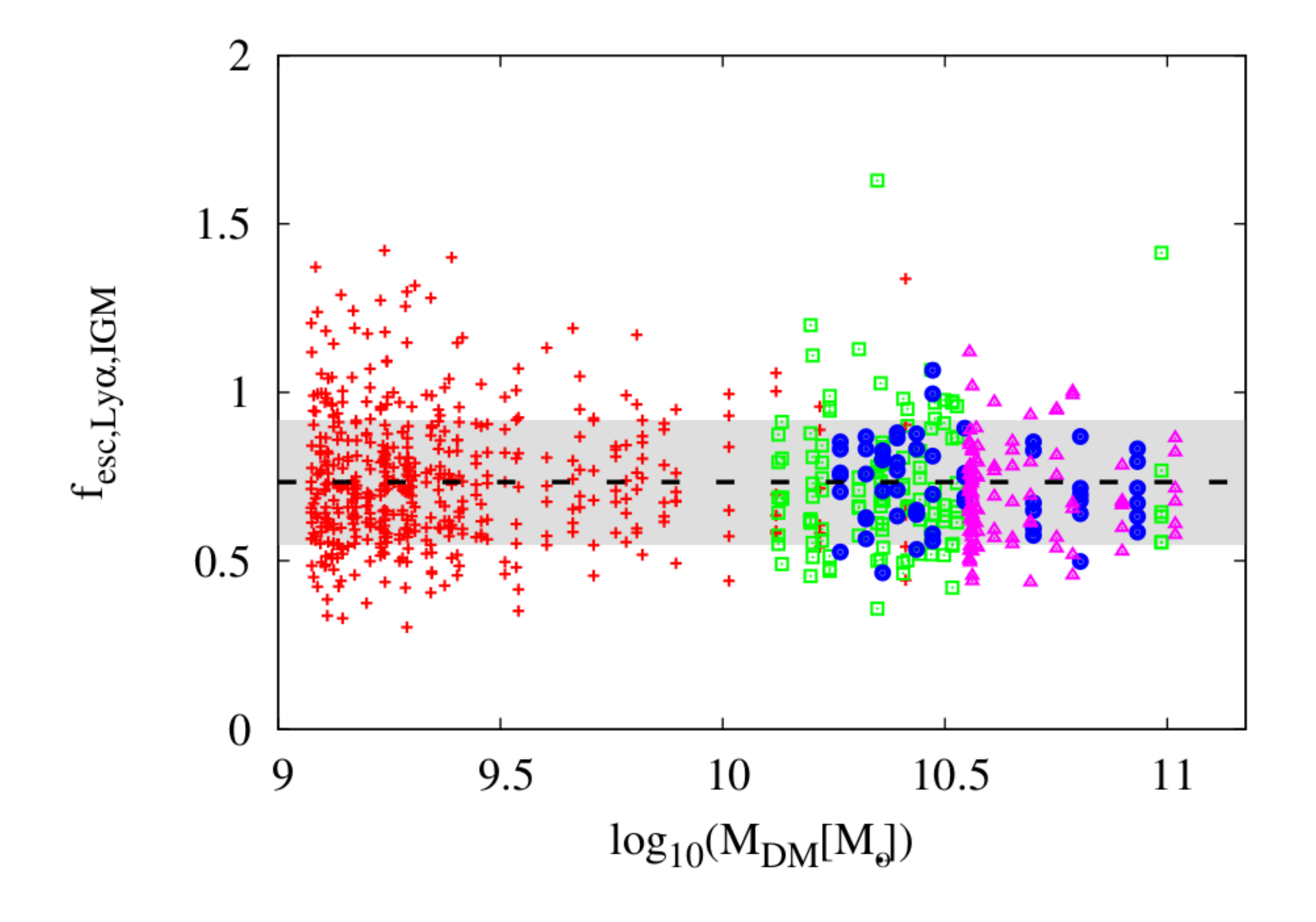}
\caption[$\fesclyaigm$ - DM mass correlation.]{Ly$\alpha$ photon escape fractions due to the IGM, $\fesclyaigm$, plotted against DM mass for objects from L05 (red crosses), L10 (green squares), L20 (blue circles)  and L30 (pink triangles) simulations. The best fit line has a log-normal slope of -0.01, with mean $\fesclyaigm$ of $\sim 0.73$ and a standard deviation $\sigma$=0.18. The $\sigma$ region is shaded in light grey.}
\label{fig:fescdm}
\end{figure}

\begin{figure*}
\centering
\includegraphics[width=130mm,height=90mm]{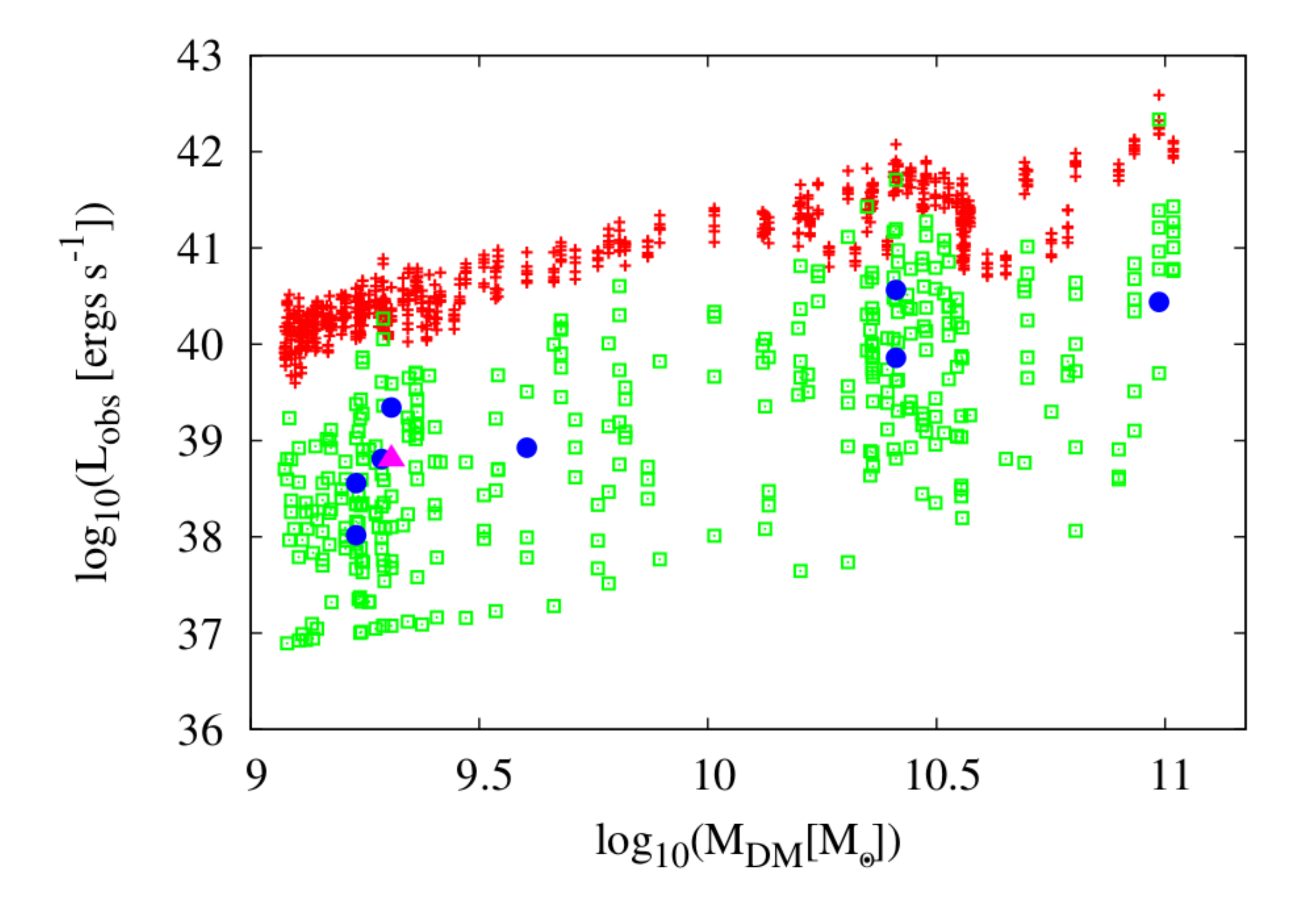}
\caption[Observed luminosity - DM mass correlations for different SB cuts.]{Observed Ly$\alpha$ luminosity of the objects from all simulations plotted against their DM mass. Symbols refer to different SB cuts, $SB_{\rm th}$: $10^{-25}$ erg~s$^{-1}$~cm$^{-2}$~arcsec$^{-2}$ (red crosses), $10^{-21}$ erg~s$^{-1}$~cm$^{-2}$~arcsec$^{-2}$ (green squares), $10^{-20}$ erg~s$^{-1}$~cm$^{-2}$~arcsec$^{-2}$ (blue circles) and $5 \times 10^{-20}$ erg~s$^{-1}$~cm$^{-2}$~arcsec$^{-2}$ (pink triangle).}
\label{fig:lumdmsb}
\end{figure*}

The detectability of the objects simulated in this work depends on the SB thresholds $SB_{\rm th}$ set by different observational campaigns. Since the lowest SB value of a pixel in the maps is $SB_{\rm min}=10^{-23}$ erg~s$^{-1}$~cm$^{-2}$~arcsec$^{-2}$, only for such a small (or lower) value of $SB_{\rm th}$ would it be possible to detect all the flux from this object. At this $SB_{\rm th}$ level the observed luminosities, calculated using Equations~\ref{eqn:fobs} and \ref{eqn:lobs}, for the source shown in the previous Figures are in the range $L_{\rm obs} = 4-12 \times 10^{41}$ erg~s$^{-1}$. For reference, the input luminosity is $L_{\rm Ly\alpha}^{\rm esc} = 9.2 \times 10^{41}$ erg~s$^{-1}$. This would lead to an escape fraction $\fesclyaigm =  L_{\rm obs}/L_{\rm Ly\alpha}^{\rm esc}$ in the range $0.43-1.3$ (see below for a discussion). 
If instead e.g. $SB_{\rm th} = 3 \times 10^{-21}$ erg~s$^{-1}$~cm$^{-2}$~arcsec$^{-2}$, only two maps would result in a detection, with luminosities $L_{\rm obs} (>SB_{\rm th}) = 8 \times 10^{40}$ erg~s$^{-1}$ and  $7.9 \times 10^{39}$ erg~s$^{-1}$, compared to $L_{\rm obs}(>SB_{\rm min}) = 1.2 \times 10^{42}$ erg~s$^{-1}$ and $8 \times 10^{41}$ erg~s$^{-1}$, respectively. Note that if observations had a threshold SB at this level only a few percent of the photons would be observed leading to a very low effective Ly$\alpha$ escape fraction $ \fesclyaigm^{\rm eff} = L_{\rm obs} (>SB_{\rm th})/L_{\rm Ly\alpha}^{\rm esc}=0.007-0.08$. For $SB_{\rm th} = 10^{-21}$ erg~s$^{-1}$~cm$^{-2}$~arcsec$^{-2}$, detections would be made in four maps with an observed luminosity in the range $L_{\rm obs} (>SB_{\rm th}) = 2 \times 10^{39} - 5 \times 10^{41}$ erg~s$^{-1}$, corresponding to $\fesclyaigm^{\rm eff} = 0.002-0.6$ for $L_{\rm Ly\alpha}^{\rm esc} = 9.2 \times 10^{41}$ erg~s$^{-1}$.

Some lines-of-sight can have an effective escape fraction $> 1$. These are lines-of-sight through voids, which in addition to the isotropic flux of photons from the central source, also get an additional contribution of photons which have been scattered off from other lines-of-sight. A Ly$\alpha$ photon encountering the surface of a high density region (like a filament close to the source) has a high probability of scattering towards the lower density gas around it. Once in a line-of-sight through a void, the photon travels longer distances without scattering while redshifting out of resonance, improving the probability of escape towards the observer. Thus lines-of-sight through voids are preferred over the ones through high density regions and get Ly$\alpha$ photon contributions from other lines-of-sight, leading to a more than average photon escape and an occasional effective escape fraction $> 1$. Note that while in the case of absorption photons removed from a line-of-sight do not contribute to any other line-of-sight leading to an escape fraction always $\leq 1$, in the case of scattering photons removed from a line-of-sight appear in another one leading to escape fractions which can be also $>1$. This makes a 3D treatment of the Ly$\alpha$ photon scattering of particular relevance for a proper LAE modelling.

\subsection{Statistical Trends}

Next we discuss how the distribution in SB affects the statistical properties of a large sample of LAEs. The difference in SB profiles in fact leads also to a spread in the Ly$\alpha$ luminosity of the same object observed from different directions, as seen earlier. The objects in our sample have an intrinsic correlation between the stellar and the DM mass, with a log-log slope of 1.12 and a standard deviation $\sigma$=0.13 dex (see Fig.~\ref{fig:stardm}). The best fit line (dashed) has been obtained using only haloes from L05 and L10. In fact, due to low particle resolution in the larger boxes, some of the haloes selected in L20 (blue circles) and L30 (pink triangles) have a stellar mass which lies below the expected trend. A correlation similar to the one shown in Figure~\ref{fig:stardm} is also expected to exist between the input Ly$\alpha$ luminosity of the source, $L_{\rm Ly\alpha}^{\rm esc}$, and the DM mass. Any additional scatter in the observed Ly$\alpha$ luminosity $L_{\rm obs}$ is instead due to the effect of the IGM.

Figure~\ref{fig:lumdm} shows $L_{\rm obs}$ of the full sample calculated for each of the six sides of the box plotted against the DM mass for each of the simulated LAEs. As in Figure~\ref{fig:stardm}, there is a strong correlation with a best fit log-log slope of 1.12, which has been calculated using only haloes from L05 and L10. The scatter now though is 0.17 dex, larger than the intrinsic one, and it is induced by the structure in the IGM.

\begin{figure}
\centering
\includegraphics[width=85mm,height=60mm]{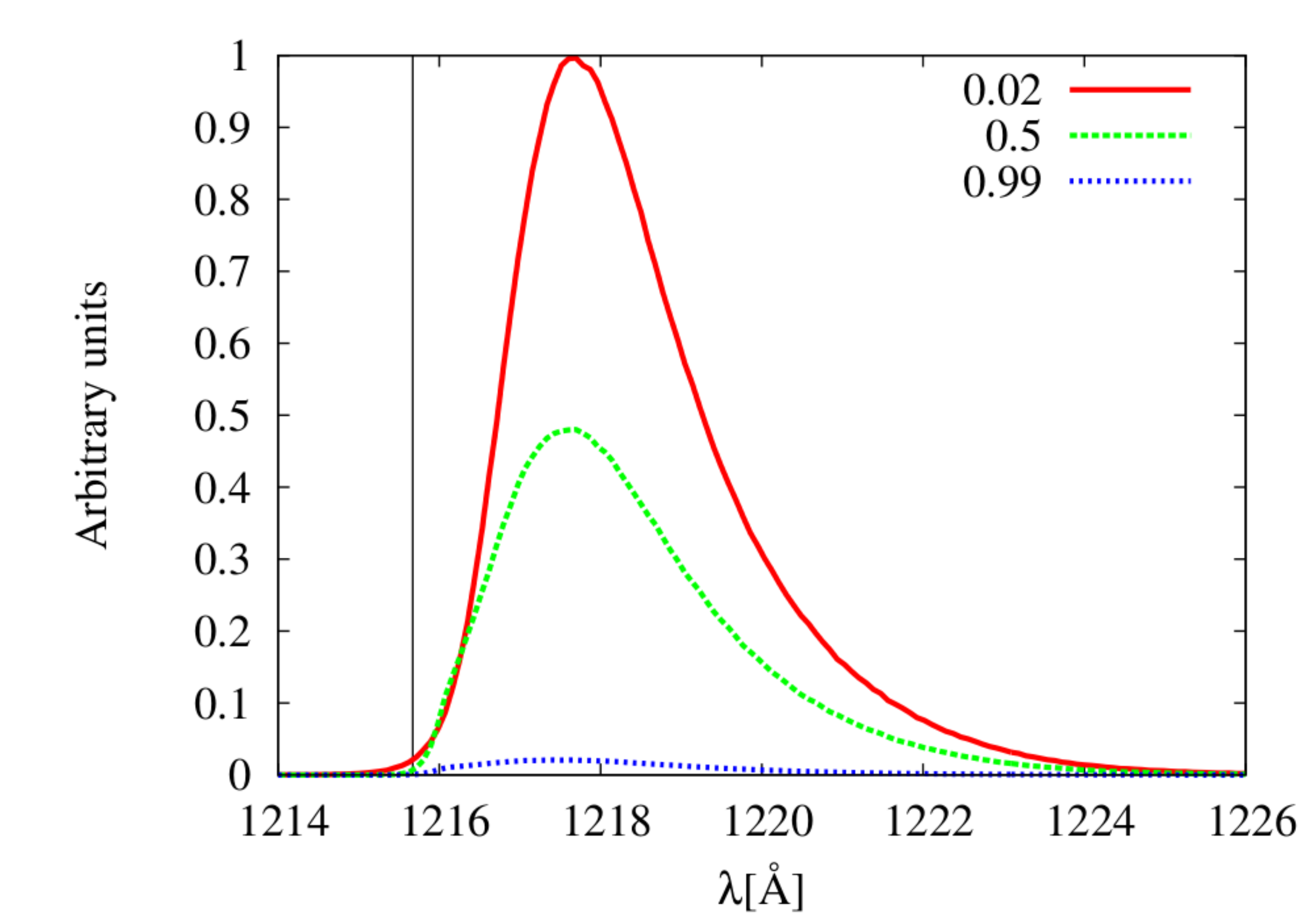}
\caption[Spectrum of $\fescionism$ test.]{Spectrum of all the Ly$\alpha$ photons exiting the simulation box for runs with different $\fescionism$. The lines refer to $\fescionism =$ 0.02 (red solid line; reference case), 0.5 (green dashed) and 0.99 (blue dotted). The vertical black line marks the Ly$\alpha$ rest-frame wavelength of $1215.67~\rm \AA$.}
\label{fig:fescion}
\end{figure}

\begin{figure*}
\centering
\includegraphics[width=150mm,height=70mm]{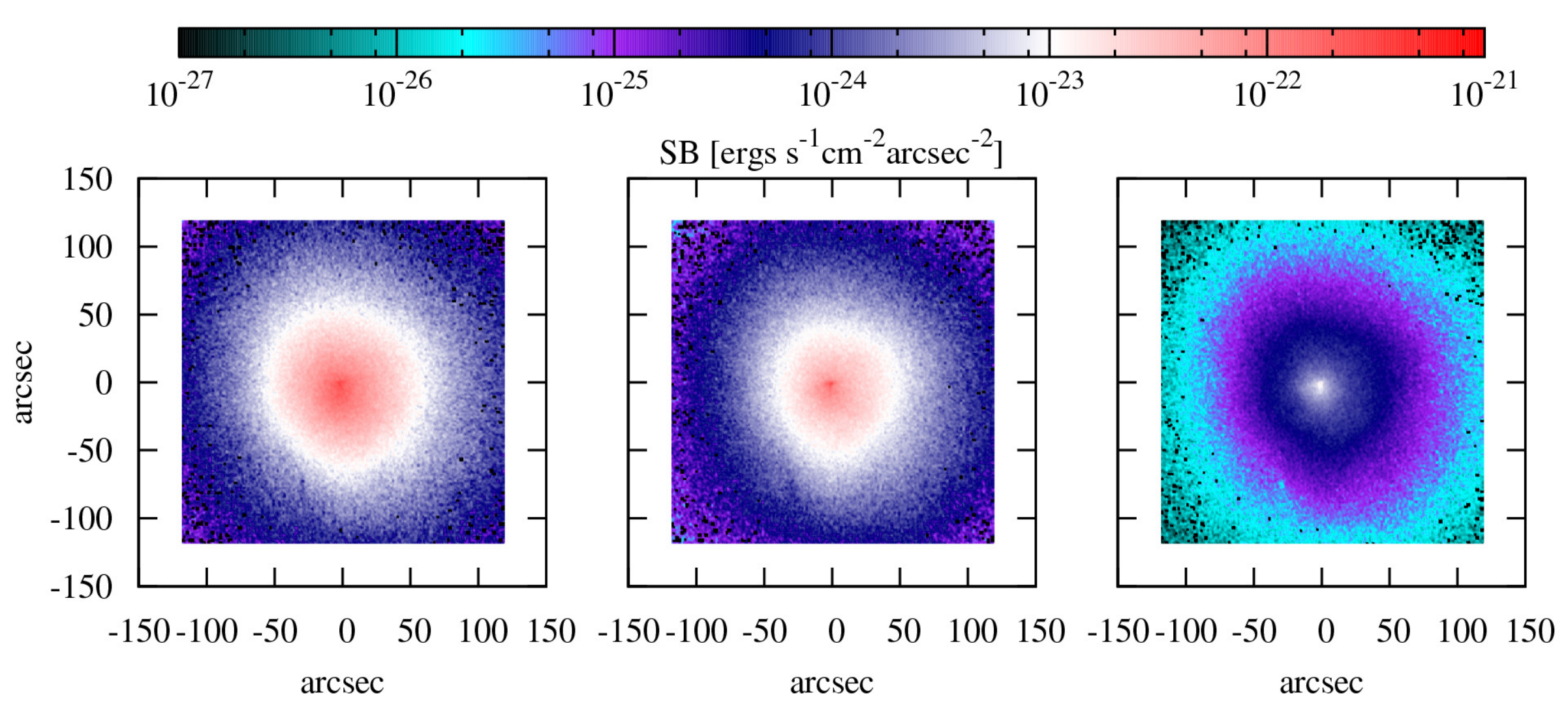}
\caption[SB maps of $\fescionism$ test.]{Surface brightness maps for one of the sides of the cube for runs with different $\fescionism$. Plotted are the maps from the simulations with $\fescionism$ = 0.02 (left; reference case), 0.5 (middle) and 0.99 (right). See text for details.}
\label{fig:fescionsb}
\end{figure*}

\begin{figure}
\centering
\includegraphics[width=85mm,height=60mm]{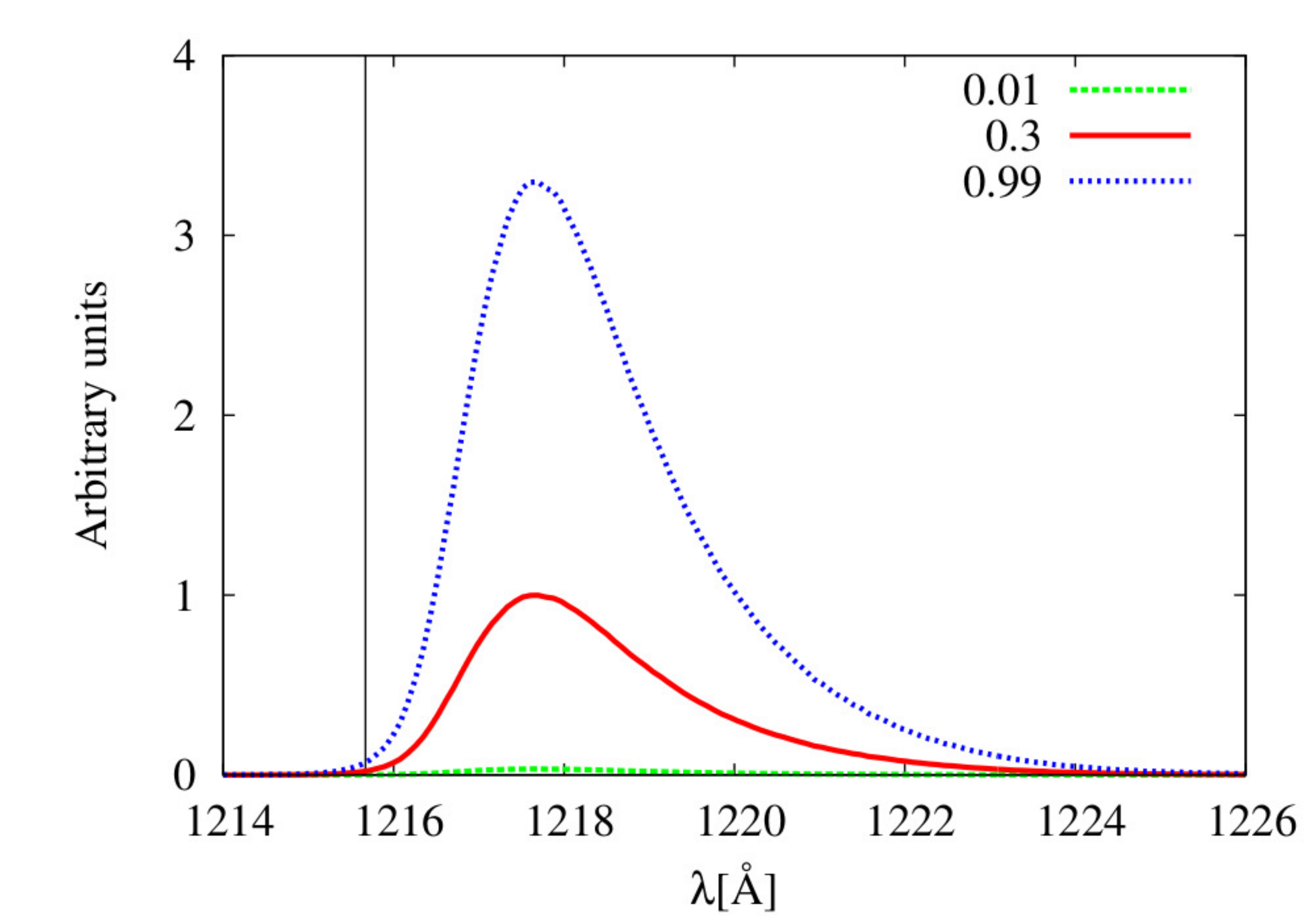}
\caption[Spectrum of $\fesclyaism$ test.]{Spectrum of all the Ly$\alpha$ photons exiting the simulation box for runs with different $\fesclyaism$. The lines refer to $\fesclyaism =$ 0.01 (green dashed line), 0.3 (red solid line; reference case) and 0.99 (blue dotted). The vertical black line marks the Ly$\alpha$ rest-frame wavelength of $1215.67~\rm \AA$.}
\label{fig:fesclya}
\end{figure}

An alternative way to look at the scatter due to the IGM is to plot the escape fraction $\fesclyaigm$ against the DM mass of the objects (Fig.~\ref{fig:fescdm}). This removes the scatter present in the intrinsic luminosity $L_{\rm Ly\alpha}^{\rm esc}$. As we can see, there is no strong correlation between $\fesclyaigm$ and the DM mass because the scatter due to the IGM structure dominates. The average escape fraction has a nearly constant value of $\sim 0.73$ with a $\sigma$ scatter of 0.18. Some lines-of-sight have escape fractions greater than 1, as explained earlier. But the statistical mean of the six observed IGM escape fractions for all the objects in the sample is 0.73. We would get an average escape fraction of 1 only if we observed all the directions for the same object and calculated the mean of all the lines-of-sight. The average of the six viewing directions is $< 1$, showing that scattering removes a large fraction of the flux for most lines-of-sight. Most of the flux from the object escapes along a small fraction of lines-of-sight through large voids, showing again the importance of IGM structure close to the source for the observability of LAEs. The spread in the correlation decreases for higher DM masses. This could be due to two reasons - a more uniform environment for haloes of higher masses leading to lesser scatter; or the resolution effects caused by the larger physical size of the grid cells ($\propto r_{\rm 200}$) for higher mass haloes compared to low mass ones. But the general consistency of $\fesclyaigm$ values in L20 and L30 with L05 seems to indicate that resolution issues might be less important. Higher resolution RT simulations of larger comoving volumes are needed to confirm this effect, which is beyond the scope of this study.

Next we investigate the detectability of this dataset by plotting in Figure~\ref{fig:lumdmsb} the observed luminosity of the objects calculated from the six sides of the simulation boxes for different SB thresholds, $SB_{\rm th}$. Only one object is detected for $SB_{\rm th} \sim 5 \times 10^{-20}$ erg~s$^{-1}$~cm$^{-2}$~arcsec$^{-2}$ (pink triangle). Note that this is not the most massive object neither in the full sample nor in the simulation box L05. Nevertheless, it looks as the brightest due to the density and velocity fields in the surrounding IGM, which lead to a very concentrated SB profile and ease the detectability of the source. From this example it is clear that the structure in the IGM plays an important role in the detectability of objects by shaping their SB profile and only a full 3D RT approach can properly capture its effect. As we lower the SB threshold, more objects are detected and a large fraction of the luminosity is accounted for. By $SB_{\rm th} = 10^{-20}$ erg~s$^{-1}$~cm$^{-2}$~arcsec$^{-2}$ (blue circles), seven objects are detected covering a mass range of two orders of magnitude. Even though there is a correlation of luminosity with DM mass apparent in the data, a large scatter is present at small masses. Decreasing the threshold by another dex, leads to the detection of almost all objects with a scatter in luminosity spanning three dex in each mass range. A SB level of $SB_{\rm th} = 10^{-25}$ erg~s$^{-1}$~cm$^{-2}$~arcsec$^{-2}$ is needed to observe the total flux from all the sources (the same as Fig.~\ref{fig:lumdm}). The SB values calculated here depend on the choice of a number of different parameters, which will be further discussed in the next Section. 

We caution the reader that current observational SB thresholds are significantly higher than the `typical' SB levels of Ly$\alpha$ radiation scattered in the IGM as found in our simulations. This leads to an effective escape fraction due to the IGM (also see Laursen et al. 2011) which is very low (a few $\%$). In addition, the observed SB is $SB_{\rm obs} \propto (1+\zz)^{-4}$ (see Eq.~\ref{eqn:sb}), making deep detections increasingly difficult at higher redshifts. Current observational thresholds for narrow band surveys are $2.64 \times 10^{-18}$ erg~s$^{-1}$~cm$^{-2}$~arcsec$^{-2}$  at $\zz=5.7$ in SXDS \citep[Subaru/XMM- Newton Deep Survey; M. Ouchi; private communication;][]{zheng10}. \cite{steidel11} achieved deep SB limits of $2.5 \times 10^{-19}$ erg~s$^{-1}$~cm$^{-2}$~arcsec$^{-2}$ for Lyman Break Galaxies (LBGs) at $\zz \sim 3$ which was obtained through stacking of 92 images. Their total observed Ly$\alpha$ flux translates to an escape fraction of Ly$\alpha$ of $\sim 0.25-0.5$ \citep[see][]{dijkstra12}. This is in agreement with Figure~\ref{fig:lumdmsb}. Obtaining these SB thresholds at higher redshifts is a very difficult task. JWST, with its predicted\footnote{From http://www.stsci.edu/jwst/science/sensitivity for $3 \sigma$ detection in $10^{5}$s with NIRSpec IFU mode at $0.1 \times 3$ arcsec$^{2}$ resolution.} $SB_{\rm th} \sim 10^{-19}$ erg~s$^{-1}$~cm$^{-2}$~arcsec$^{-2}$, would be an important step for the deep detections of LAEs.

\section{Parameter Study}
\label{sec:paramstudy}

In this Section we study the dependence of our results on the different parameters adopted. To this aim we use the biggest object in L05 with a dark matter mass of $2.5 \times 10^{10}~\msun$. The reference values are those described in Section~\ref{sec:rtsim}. 

A minor contribution to the observed flux comes from the photons emitted by the recombining gas in the IGM, which has been estimated to be $\sim 3\%$. This is not significant compared to the flux from the central source, but it should be kept in mind that this flux is not affected by the absorption in the ISM and thus has a higher escape probability. Since it is not concentrated as a point source, its contribution to the detectability of the object is negligible; rather, it might reduce the strength of the point source by spreading the flux to the scattered component of the angular directional distribution. We do not include the re-emission photons in our parameter study.

\subsection{Escape Fraction}
\label{sec:escfrac}

\begin{figure}
\centering
\includegraphics[width=85mm,height=60mm]{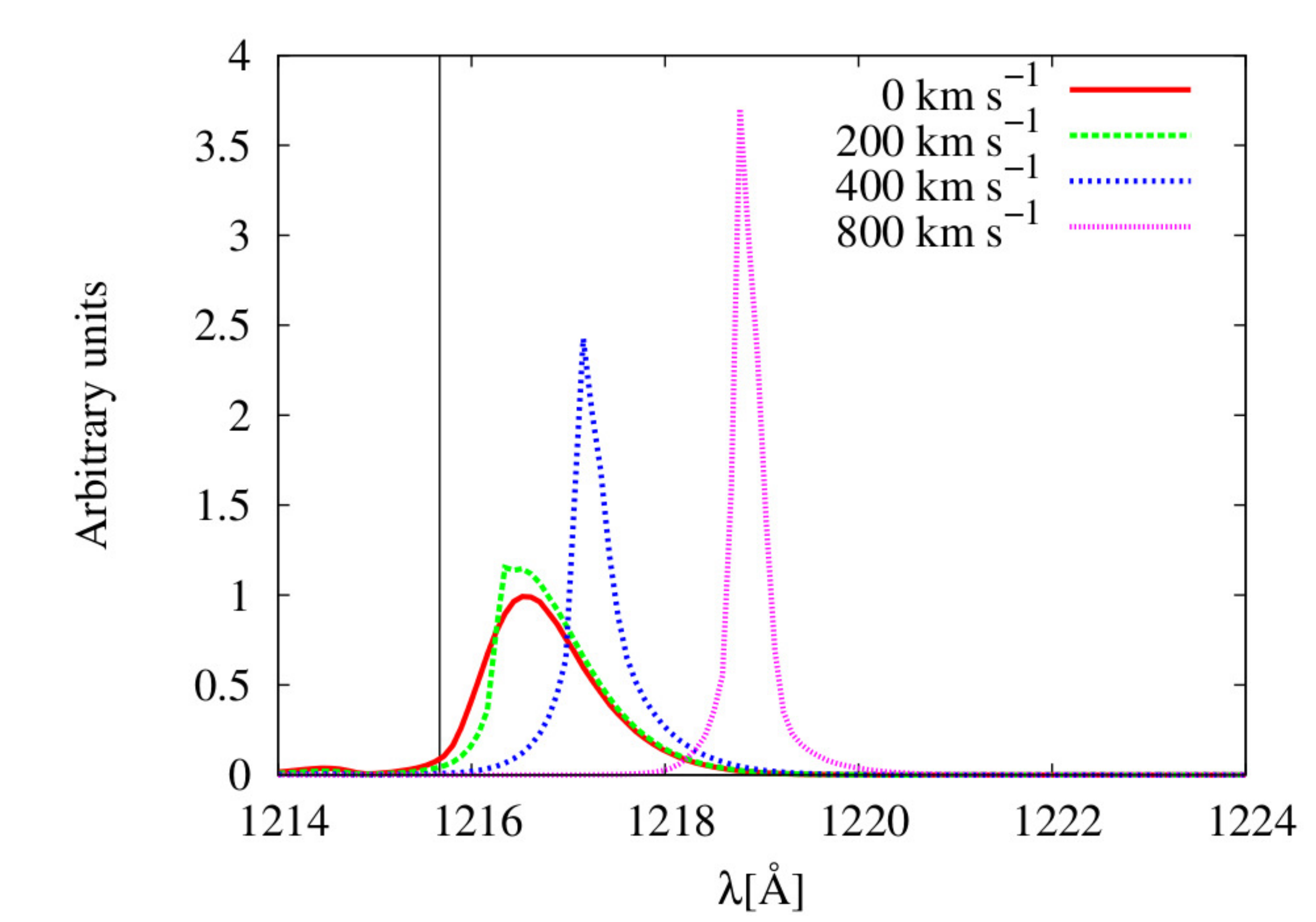}
\caption[Spectra for initial velocity test.]{Spectrum of all the Ly$\alpha$ photons exiting the simulation box for different velocity shift of the input Ly$\alpha$ monochromatic spectrum. The lines refer to a velocity shift of 0 (red solid; reference case), 200 (green dashed), 400 (blue short dashed) and 800 (pink dotted) km~s$^{-1}$.}
\label{fig:shiftspec}
\end{figure}

As already discussed, we need to specify the value of both ionizing, $\fescionism$, and Ly$\alpha$, $\fesclyaism$, escape fractions from the ISM. The amount of ionizing photons reaching the IGM is controlled by $\fescionism$, thus determining the extent of the ionized region through which the Ly$\alpha$ photons propagate. The value of the escape fraction also affects the production of Ly$\alpha$ photons by recombinations in the ISM. In our reference simulations, the box around the object has a size of $\sim 35~r_{\rm 200}$ comoving. Because this is not large enough to contain the HII region produced by $\fescionism = 0.99$, for this parameter study we use an object with DM mass of $2.29 \times 10^{10}~\msun$ from L10. We extract a cube with a side of $\sim 117~r_{\rm 200}$ comoving, which is gridded to a 256$^{3}$ grid.

To quantify the effect of changing the value of $\fescionism$ on the spectrum, in Figure~\ref{fig:fescion} we show the Ly$\alpha$ spectrum of the photons exiting the box for simulations with $\fescionism = 0.02, 0.5, 0.99$, while $\fesclyaism$ is set to the default value of $0.3$. Note that the photon packets are always blueshifted back to the source position from the edge of the box. Also note that the spectrum is for all the photons exiting the box throughout the simulation time. The Ly$\alpha$ photon count increases with decreasing $\fescionism$ (see Eq.~\ref{eqn:lyaism}), as recombination in the ISM is the dominant contribution to Ly$\alpha$ photon production. It also affects the shape of the spectrum by moving its peak closer to $1215.67~\rm\AA$, as the photons escape with less scatter. This can also be seen from the SB maps in Figure~\ref{fig:fescionsb}. The higher the $\fescionism$ is, the more concentrated is the SB profile. But due to the lower production of Ly$\alpha$ photons in the ISM, the SB values in each pixel decrease for higher $\fescionism$. Thus, increasing $\fescionism$ reduces the overall detectability of the source at a specific $SB_{\rm th}$, but improves its detectability at lower $SB_{\rm th}$ with respect to a dimmer galaxy with lower $\fescionism$ giving similar total Ly$\alpha$ fluxes.

Direct estimation of $\fescionism$ from observations is a very difficult task \citep[e.g.][]{bland01, steidel01, shapley06, iwata09, siana10}. A large effort has gone into modelling the correlation between halo properties and ionizing escape fraction from the galaxy \citep[e.g. ][]{dove00,wood00,ricotti00, ciardi02,clarke02,fujita03,razoumov06, gnedin08,wise09, yajima09,yajima11a,paardekooper11,yajima11b}. Estimated escape fraction values range from 10$^{-5}$ to 1 depending, among others, on the mass, redshift, dust and gas distribution and star formation rate, with different studies giving different values. Since the correlation between halo properties and $\fescionism$ is not well known, we choose our default value to be $\fescionism=0.02$, as in \cite{gnedin08}.

\begin{figure*}
\centering
\includegraphics[width=100mm,height=120mm]{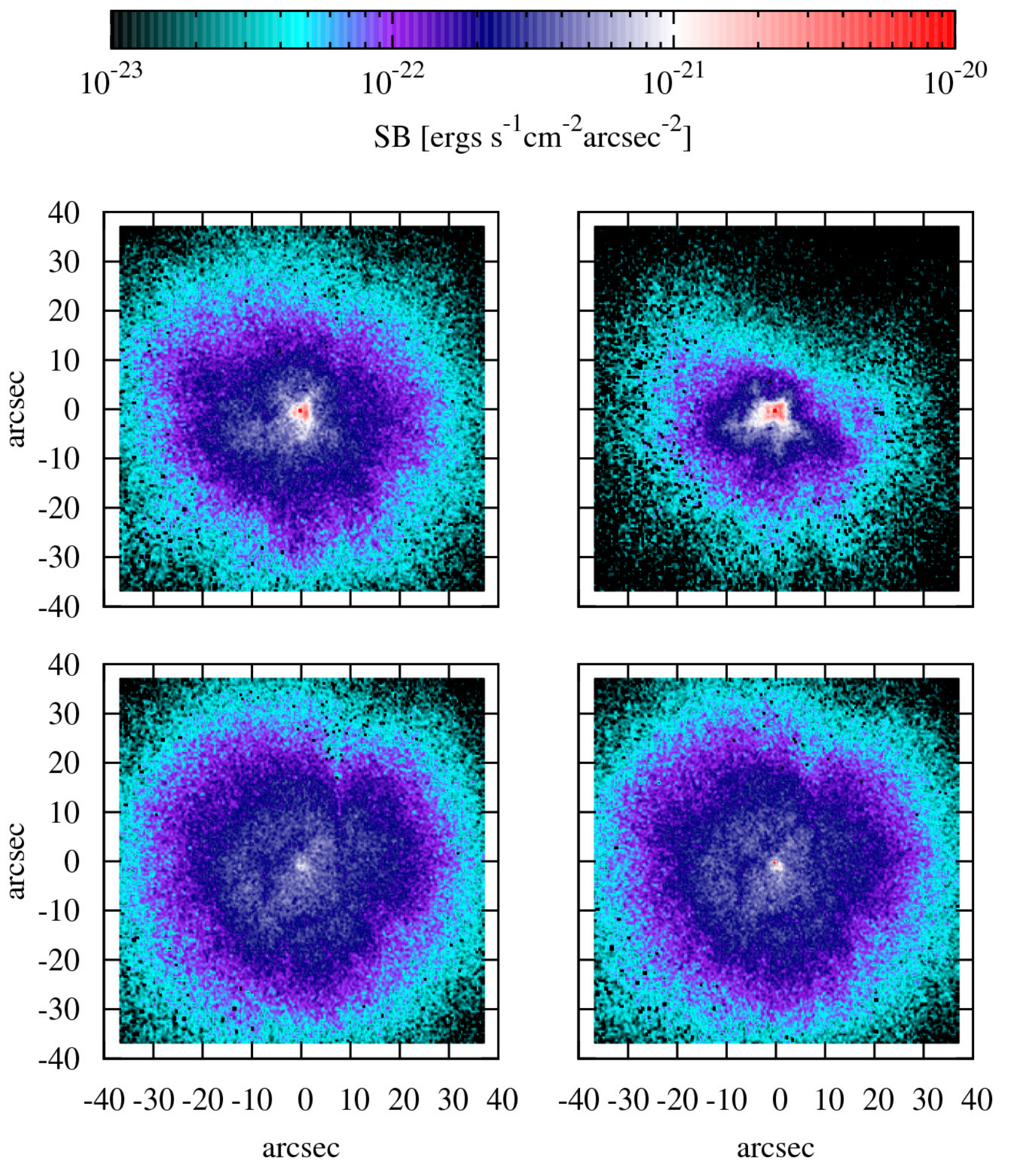}
\caption[SB maps for initial velocity test.]{Surface brightness maps for one of the sides of the simulation box and for different velocity shift of the input Ly$\alpha$ monochromatic spectrum, i.e. 0 (bottom left panel), 200 (bottom right), 400 (top  left) and 800 (top right) km~s$^{-1}$.}
\label{fig:shiftspecsb}
\end{figure*}

\begin{figure}
\centering
\includegraphics[width=85mm,height=60mm]{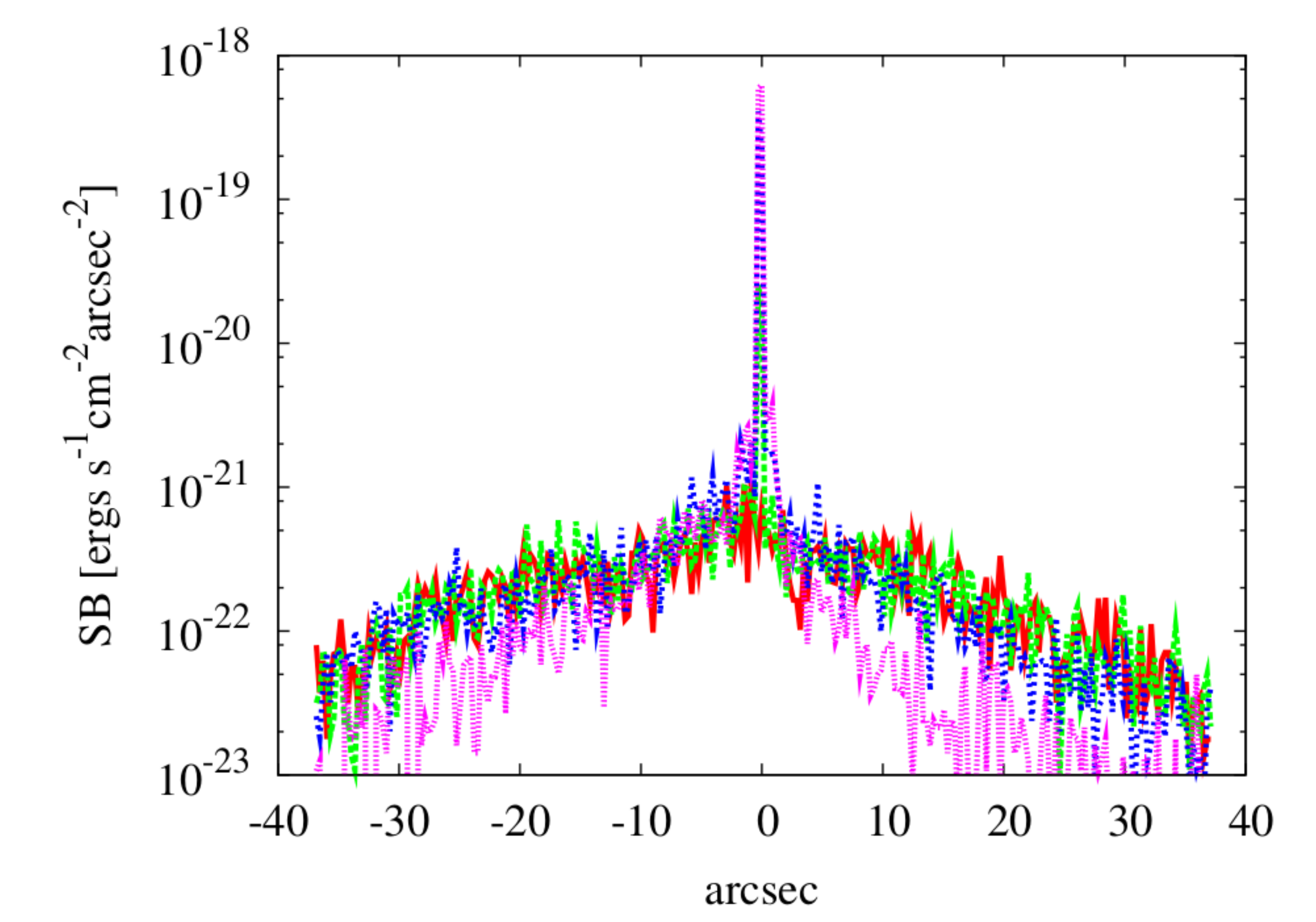}
\caption[Cross section of the SB maps for initial velocity test.]{Cross section of the SB profile for the maps in Figure~\ref{fig:shiftspecsb}. Plotted are the SB values along the y axis for an x axis value of 0. The lines refer to a velocity shift of the input Ly$\alpha$ monochromatic spectrum of 0 km~s$^{-1}$ (red solid; reference case), 200 km~s$^{-1}$ (green dashed), 400 km~s$^{-1}$ (blue short dashed) and 800 km~s$^{-1}$ (pink dotted).}
\label{fig:shiftspecsbprofile}
\end{figure}

We have then tested the effect of the escape fraction of Ly$\alpha$ photons due to the ISM, $\fesclyaism$, which controls the fraction of Ly$\alpha$ photons reaching the IGM, as detailed in Equation~\ref{eqn:dust}. We use the same object and box of the previous tests, with $\fesclyaism=0.01, 0.3$ and $0.99$. Figure \ref{fig:fesclya} shows the spectrum of the photons exiting the cube, whose intensity increases with increasing $\fesclyaism$ as more photons reach the IGM. On the other hand, the shape of the spectrum is not altered. Similarly, the morphology of the SB profile is not affected.

The escape fraction of Ly$\alpha$ photons from the ISM has been estimated in a large number of studies \citep[e.g.][]{ledelliou05, ledelliou06, dave06, nagamine10, ouchi08, dayal09, kobayashi07, laursen09, steidel11, schaerer11, forero11, yajima11b}. Since the amount and type of dust in the ISM of a galaxy and its correlation with other halo properties is highly unknown, we choose a reference value of $0.3$. This adopted value is similar to the value that was required by \cite{dayal09} in order to reproduce the observed Ly$\alpha$ luminosity functions at $z=5.7$ (note that this requirement assumes that the IGM transmits $\sim 50\%$ of all photons: for a more transparent/opaque IGM, the required escape fraction would be lower/higher).

\begin{figure*}\centering\includegraphics[width=180mm,height=60mm]{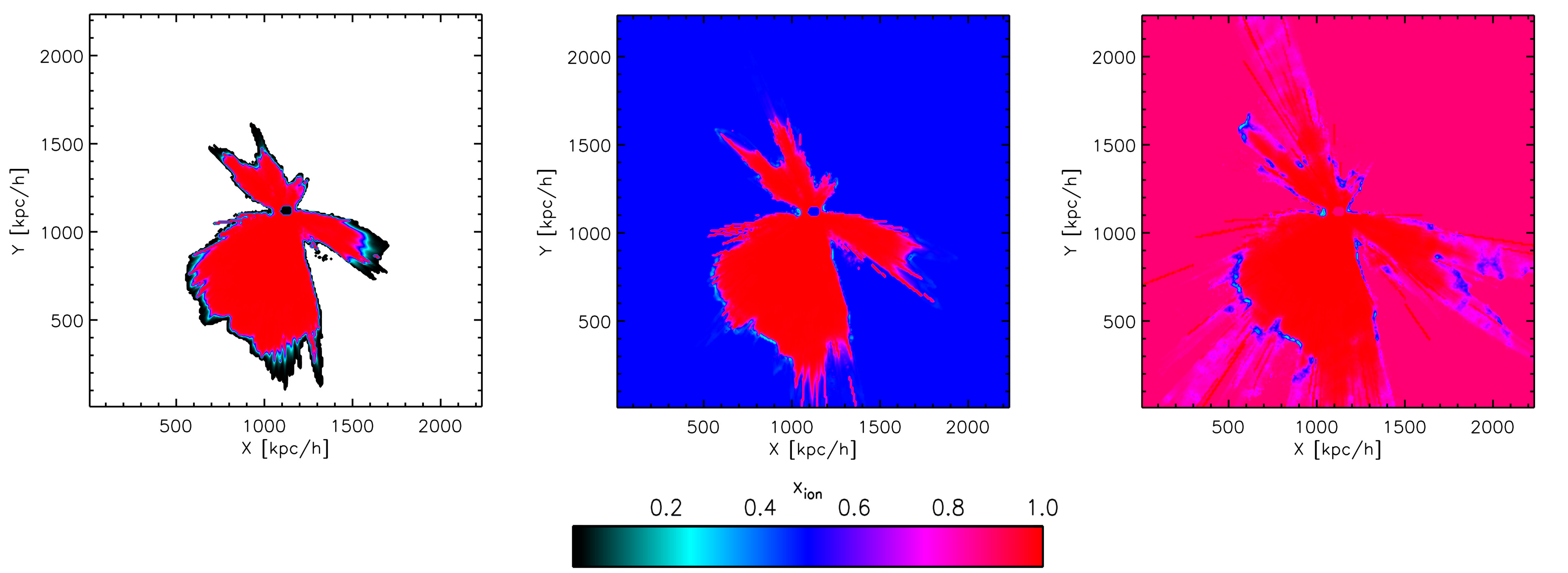}
\caption[Slice through the ionization structure for initial ionization level test.]{Slice through the final ionization structure for simulations with initial volume averaged ionization fractions $\ionfrac$=0 (left panel; reference case), 0.5 (middle) and 0.89 (right).}
\label{fig:ionfracstruct}
\end{figure*}

\begin{figure}\centering\includegraphics[width=85mm,height=60mm]{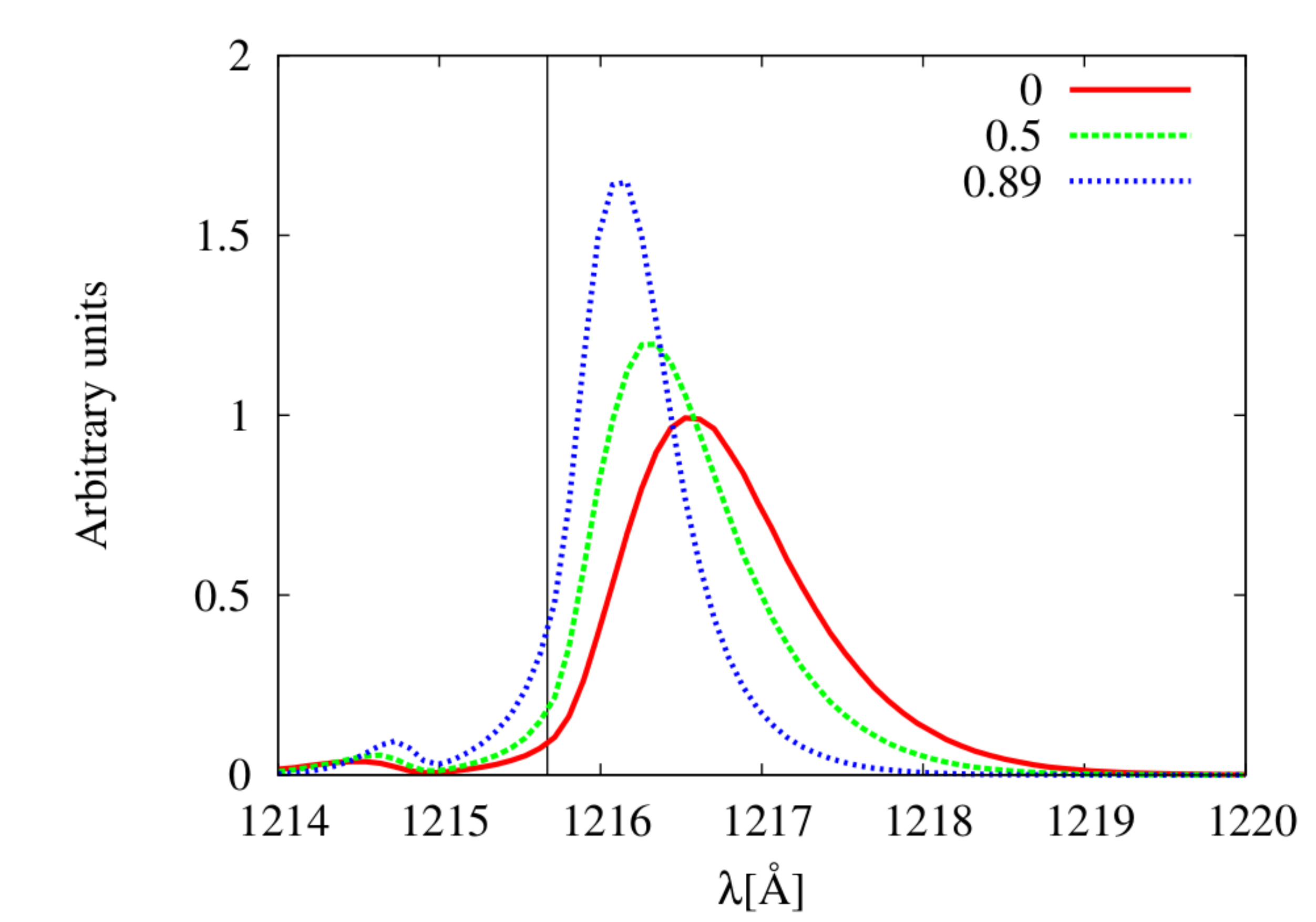}
\caption[Spectra for initial ionization level test.]{Spectrum of all the Ly$\alpha$ photons exiting the simulation box for runs with initial volume averaged ionization fraction $\ionfrac$=0 (red solid line; reference case), 0.5 (green dashed) and 0.89 (blue dotted).}
\label{fig:ionfrac}
\end{figure}

\subsection{Effect of Input Ly$\alpha$ Spectrum}
\label{sec:velshift}

The RT of Ly$\alpha$ photons through the IGM is affected by their wavelength as they escape the ISM. In our default runs, we use a monochromatic line at $1215.67~\rm\AA$, but we can expect the spectral shape of the photons emitted by the stellar sources to be distorted by their interaction with the ISM before entering the IGM.  In fact, observations of LAEs at $\zz\sim2-3$ show a non-monochromatic spectrum \citep{verhamme07}. Other authors \citep[e.g.][]{dayal09,zheng10,dijkstra10} use gaussian profiles with a width given by a fraction of the circular velocity at the virial radius of the DM halo, with wider profiles leading to an easier escape of the Ly$\alpha$ photons from the surrounding medium. A lot of effort has gone into modelling more accurately the Ly$\alpha$ line profile for a range of parameters  \citep{schaerer11}, but the exact shape of the spectrum and its connection to the galaxy properties are still unclear. Because of all these uncertainties, we choose to investigate the effect of the input Ly$\alpha$ spectrum simply by shifting a  monochromatic spectrum.

\begin{figure*}\centering\includegraphics[width=150mm,height=70mm]{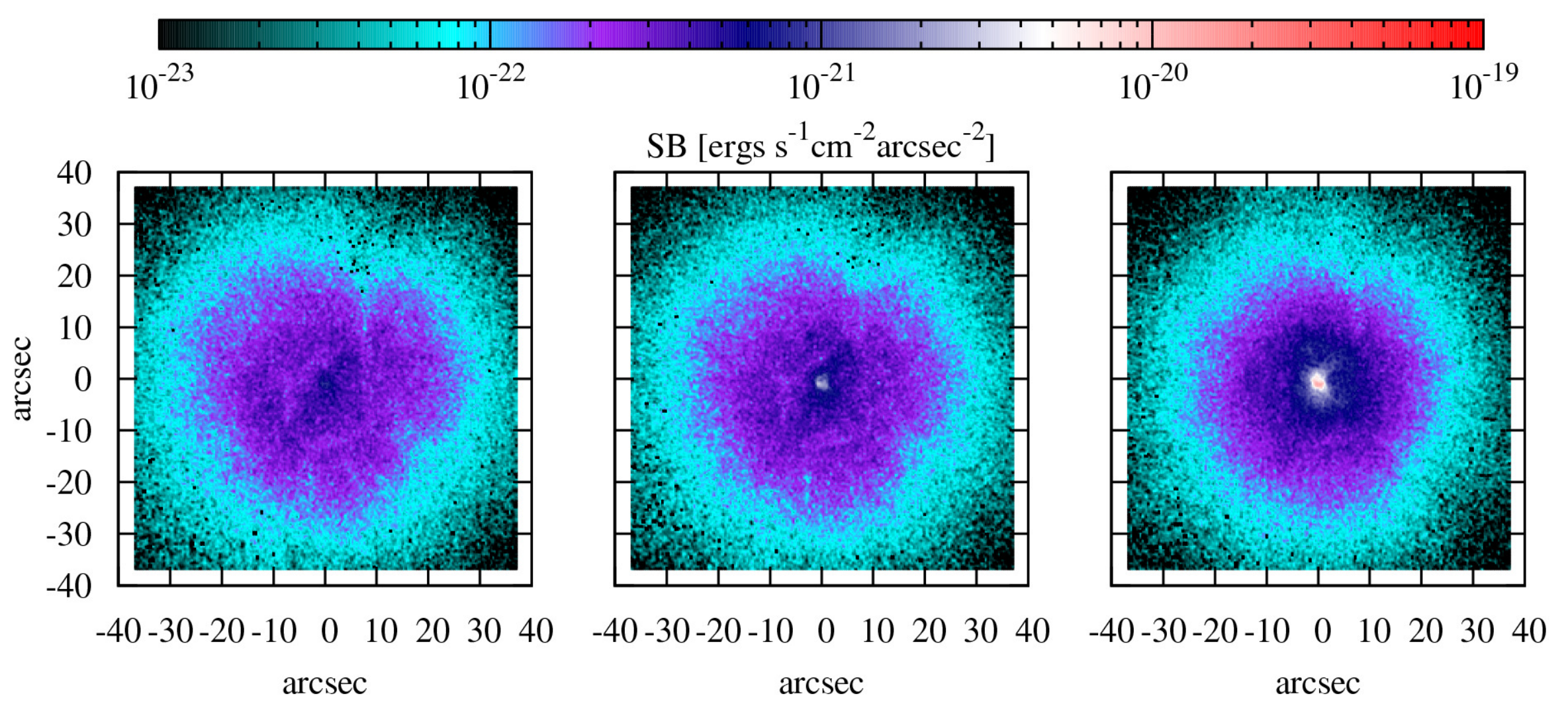}
\caption[SB maps for initial ionization level test.]{Surface brightness maps for one of the sides of the box for runs with initial volume averaged ionization fraction $\ionfrac$= 0 (left panel; reference case), 0.5 (middle) and 0.89 (right).}
\label{fig:ionfracsb}
\end{figure*}

\begin{figure}
\centering
\includegraphics[width=85mm,height=60mm]{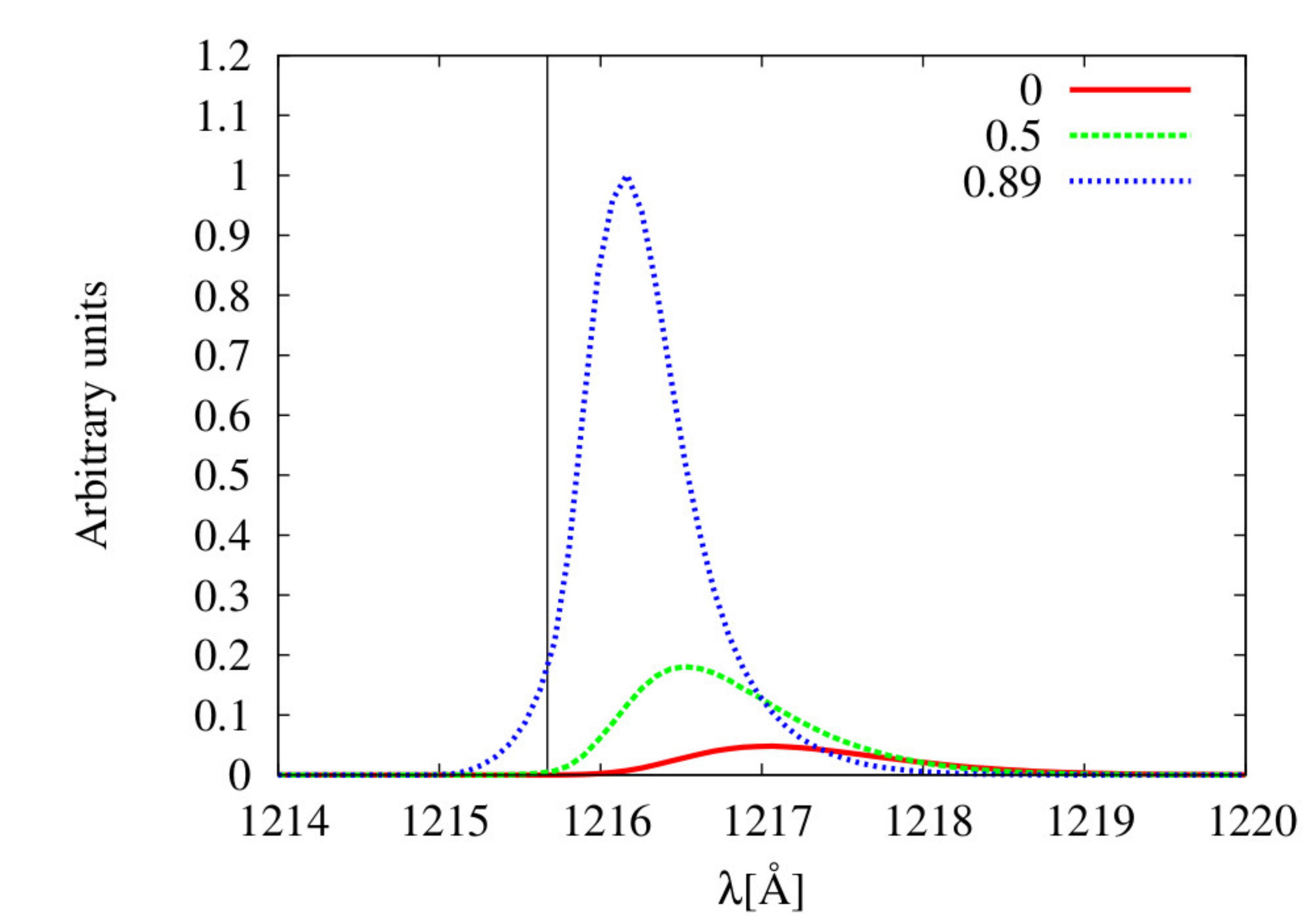}
\caption[Spectra for initial ionization level test after including scattering beyond the RT simulation cube.]{Same as Figure~\ref{fig:ionfrac} but including scattering of Ly$\alpha$ photons by the column of neutral gas beyond the edge of the cube. The y-axis has the same scale as  Figure~\ref{fig:ionfrac}. The lines are for a volume averaged ionization fraction $\ionfrac$=0 (red solid line; reference case), 0.5 (green dashed) and 0.89 (blue dotted).}
\label{fig:suppxion}
\end{figure}

Figure~\ref{fig:shiftspec} shows the spectrum of all the Ly$\alpha$ photons exiting the box for different values of the input Ly$\alpha$ monochromatic photons, i.e. with velocity shifts of 0 (reference case), 200, 400 and 800 km~s$^{-1}$. Since the scattering cross section is $\propto \Delta\nu^{-2}$, the larger the shift, the lesser the scatter suffered by the photons. The velocity field in the IGM also plays a role in determining the scattering probability of these photons. Since the velocity distribution of the IGM around the object has a complex structure, it is not possible to uniquely predict the velocity shift at which the photons stop being scattered. For our test object, as the photons are shifted by 200 km~s$^{-1}$, the scattering is reduced by a factor of two, which in turn is reflected in a reduction of photons at $1215.67~\rm\AA$. For a shift of 400 km~s$^{-1}$, the scattering is reduced by an order of magnitude. Monochromatic spectra shifted by $\ge$ 800 km~s$^{-1}$ get hardly scattered. This also affects the SB profiles, as shown in Figure~\ref{fig:shiftspecsb}, where the SB in the four cases is plotted for one of the sides of the box (top-left map in Figure~\ref{fig:run1sb}). As the velocity shift increases, the probability of scatter of photons decreases, leading to a more concentrated SB profile. In addition, only a small fraction of the photons are scattered, forming a low SB halo. This can be seen more clearly in Figure~\ref{fig:shiftspecsbprofile} where the cross section of the maps is shown. The larger the velocity shift is, the higher is the SB value in the central pixel, improving the detectability of the source. Also, a larger fraction of the flux escapes from this pixel. For example, 0.004, 0.1, 1.5 and 4 per cent of the total flux escapes
through this pixel for the cases of no velocity shift, a velocity shift of 200, 400 and 800 km~s$^{-1}$ , respectively. Therefore, the initial spectra of Ly$\alpha$ photons (i.e. the shape induced by the processing of the Ly$\alpha$ photons in the ISM) has a very important effect in determining the SB profile of the object and on the observability of LAEs. Quantifying this in more details is beyond the scope of this study and will be investigated elsewhere.

\begin{figure*}
\centering
\includegraphics[width=180mm,height=60mm]{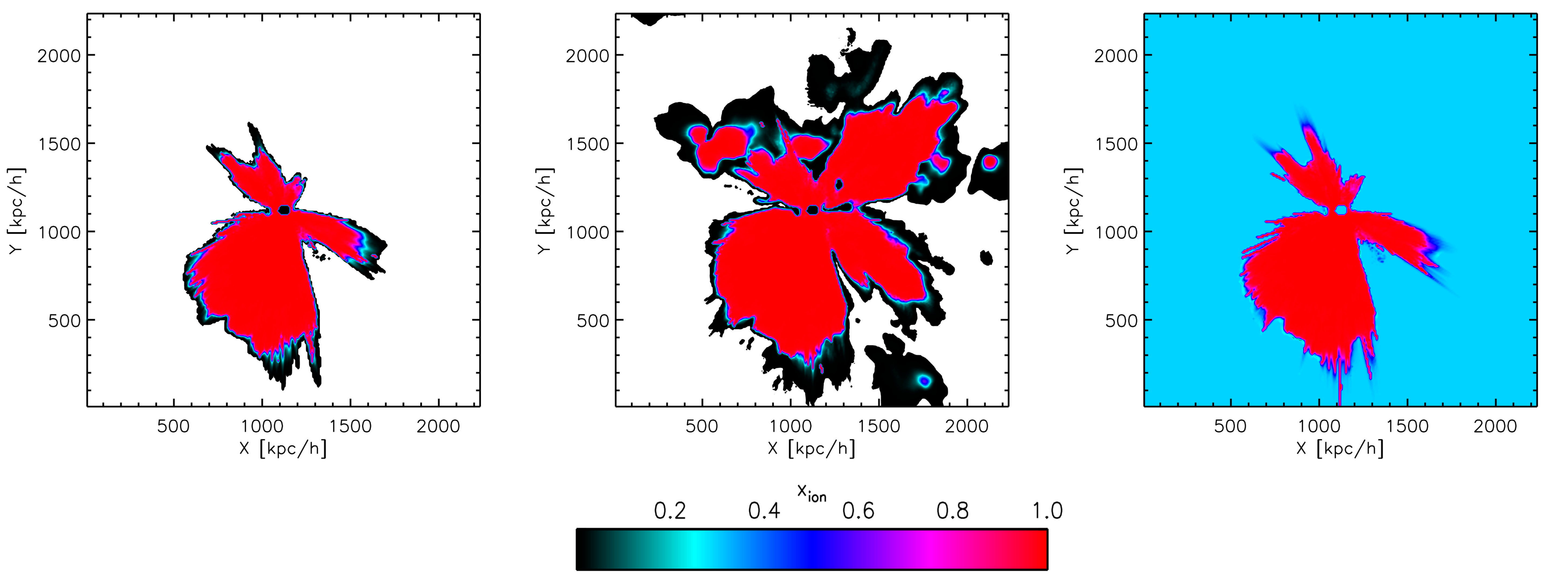}
\caption[Slice through the ionization structures for clustering test.]{A slice through the ionization structure for RT simulations showing the effect of clustering of sources. The ionized region is produced by the central source alone (left panel), all the sources in the simulation box (case $\mathcal C$, middle panel) and the central source alone in a uniform partially ionized IGM (case $\mathcal U$, right panel). The final volume averaged ionization fraction in the middle and right panel is the same. See text for further details.}
\label{fig:clusterion}
\end{figure*}

\begin{figure*}
\centering
\includegraphics[width=150mm,height=70mm]{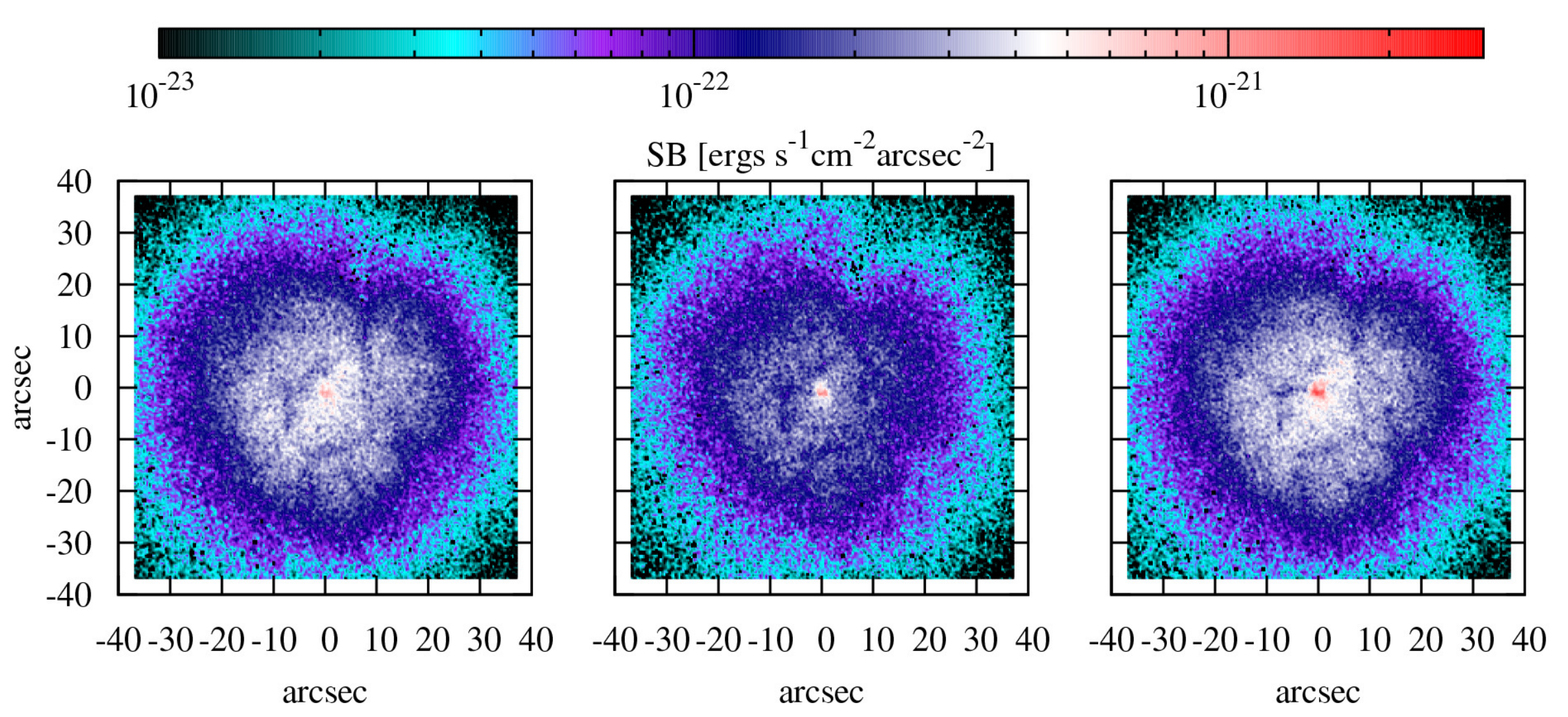}
\caption[SB maps for clustering test.]{Surface brightness maps for one of the sides of the simulation box for the same simulations of Figure~\ref{fig:clusterion}.}
\label{fig:clustersb}
\end{figure*}

From Ly$\alpha$ spectrum models of \cite{schaerer11}, the peak of the spectra of photons escaping from the galaxy is at wavelengths where the velocity shift is equivalent to $\sim 1.5-2$ times the outflow velocities. The typical outflow velocities estimated from observations by \cite{verhamme07} is $\sim 150-200$ km~s$^{-1}$, thus showing a Ly$\alpha$ spectrum peak at $200-400$ km~s$^{-1}$. With a simplifying assumption that almost all the flux enters the IGM at these wavelength, we can estimate the expected changes in observed luminosity\footnote{Note that a specific SB threshold might select more than 1 pixel.} at different $SB_{\rm th}$ due to this velocity shift. At $SB_{\rm th} = 10^{-21}$ erg~s$^{-1}$~cm$^{-2}$~arcsec$^{-2}$, for a velocity shift of 200 (400) km~s$^{-1}$,  $\fesclyaigm^{\rm eff} = 0.01$ (0.1), compared to  $7 \times 10^{-4}$ in our reference case with no velocity shift. Thus detectability of a LAE at a specific $SB_{\rm th}$ can improve by orders of magnitude for realistic velocity shifts of $200-400$ km~s$^{-1}$ even with $\ionfrac = 0$ \citep[see also][]{dijkstra10}. Therefore, understanding the effect of the ISM of the galaxies along with their outflows is a crucial input for better modelling of LAEs. This topic will be addressed elsewhere in further detail.

\subsection{Effect of IGM Ionization}
\label{sec:ionization}

Another important factor determining the observability of LAEs is the ionization level of the IGM outside the fully ionized region created by the source. The initial ionization fraction of our simulations is set to a uniform default value of $\ionfrac= 0$ for convenience, but simulations of the reionization process indicate that most of the IGM is substantially ionized at the redshifts studied in this paper \citep[e.g.][]{ciardi11}. The IGM ionization level is determined by a combination of a background radiation and local sources. Here we study how these two contributions change the observability of our test source.

\subsubsection{Uniform Ionization Distribution}
\label{sec:unifion}

To understand the effect of the IGM ionization level on the Ly$\alpha$ RT, we repeat the reference simulation with an initial volume averaged ionization fraction of $\ionfrac=0.5$ and 0.89. The first value corresponds to an epoch when only half of the IGM is ionized, while the second is the value obtained from the reionization model ${\mathcal E}$1.2-$\alpha$1.8 described in \cite{ciardi11}, which is designed to be consistent with existing observational constraints such as the Thomson scattering optical depth and the photo-ionization rate at $z<6$.

Figure~\ref{fig:ionfracstruct} shows a slice cut through the ionization structure at the end of the simulation for the three cases described above. As expected, the higher $\ionfrac$ is, the larger is the fully ionized region produced by the source, although the differences are not dramatic because the high density filaments surrounding the source effectively confine the fully ionized region in all cases and determine its shape. Note that while the edges of the ionized region are controlled by the high density filaments in the IGM, the inclusion of ionizing flux from galaxies in those high density clumps could lead to a different ionization structure (as shown in Sec.~\ref{sec:cluster}).

As a larger ionized region results in less scatter for the Ly$\alpha$ photons, we expect an easier escape in this case. This is shown in Figure~\ref{fig:ionfrac}, where the spectrum of all the Ly$\alpha$ photons exiting 
the simulation box is plotted. Due to the lower number of scatterings undergone by the Ly$\alpha$ photons for higher values of $\ionfrac$, the peak shifts to bluer wavelengths and the photon count at the peak increases while reducing the width of the line profile. The SB maps for one of the sides of the box for the different $\ionfrac$ values is shown in Figure \ref{fig:ionfracsb}. As $\ionfrac$ increases, the SB profile becomes more concentrated while improving detectability. From the maps, we can infer that observations at $SB_{\rm th} = 3 \times 10^{-21}$ erg~s$^{-1}$~cm$^{-2}$~arcsec$^{-2}$ would appear as a point source to the observer. For this viewing direction, the luminosity (effective escape fraction) of the point source detected for $\ionfrac=0.5$ and 0.89 is $8.4 \times 10^{38}$ erg~s$^{-1}$ (0.001) and $3.7 \times 10^{40}$ erg~s$^{-1}$ (0.04), respectively, compared to a non detection at $\ionfrac=0$. More specifically, if we were to consider the maps from all six faces of the cube, at $\ionfrac=0.89$ the object would be detected as a point source from all the six directions  with $\fesclyaigm^{\rm eff}$  in the range $0.004-0.6$, compared to only two detections as point source in six maps at $\ionfrac=0$ with $\fesclyaigm^{\rm eff} = 0.008$ and 0.09. Thus at high ionization fractions most of the objects can be detected as point sources at lower SB thresholds compared to our reference case. Our results appear similar to \cite{zheng10} who find that even at a mean neutral fraction of 10$^{-4}$, only a small fraction of the photons (8-33 $\%$ of the intrinsic luminosity) escape as a point source.

Because this paper is primarily focused on the effect of the IGM in the immediate surroundings of a source of Ly$\alpha$ photons, we do not follow the propagation of the photons to the observer. To do so, we would need much larger boxes, losing the resolution necessary to properly account for the scales we are mainly interested in here. Nevertheless, the IGM outside our simulation box is bound to scatter the photons in an amount dependent on their frequency when they escape the box. The optical depth $\tau$ due to scattering away from the line-of-sight to the observer (at $\zz = 0$) for a LAE at $\zz=\zz_{s}$ is given in Equations~\ref{eqn:tau1} and~\ref{eqn:tau2}. 

For $x_{\rm HI}=1$ at $\zz_{s} = 7.7$, the optical depth $\tau\sim1$ for $ \Delta v \sim 1400 \rm\; km\;s^{-1}$, corresponding to a comoving distance of $8.6~\Mpch$ from the source which is much larger than our cube sizes. Beyond this distance from the source the Ly$\alpha$ photons have redshifted far enough from resonance such that the neutral IGM does not further affect the Ly$\alpha$ SB profiles. However, in some cases analyzed in this work -- in particular low mass objects for which smaller boxes are extracted -- the photons emitted by the source do not redshift far from the line resonance when they reach the edge of the box. The neutral gas outside the box can then further modify the predicted Ly$\alpha$ SB profile. 

As a simple estimate of the scattering in neutral gas beyond the boundaries of the cube we can use Equation~\ref{eqn:tau2}. The effect of this scattering on the spectra of photons escaping the box for our test object is shown in Figure~\ref{fig:suppxion}. 
For $\ionfrac=0$, the $\tau$ integrated over the whole spectrum is 2.9, but as the ionization increases, the $\tau$ decreases to 1.8 at $\ionfrac=0.5$ and to 0.55 at $\ionfrac=0.89$, leading to more photons being observed. The scattering beyond the cube can therefore be significant for small $\ionfrac$, resulting in lower SB values per pixel extending over a larger angular size. Even deeper SB cuts would then be required to observe the total luminosity of the object. In contrast, for high $\ionfrac$ scattering beyond the cube can be safely ignored, and the SB profile would retain the point source like appearance seen in Figure~\ref{fig:ionfracsb}. The qualitative results described in the previous sections (e.g. a spread and increment in $\fesclyaigm^{eff}$ for lower $SB_{\rm th}$) would still hold.  Another factor which reduces scattering outside the simulation box is the velocity shift discussed in Section~\ref{sec:velshift}. A shift of 200 (400) km~s$^{-1}$ reduces the effective $\tau$ to 2.8 (2.3) even for $\ionfrac=0$. This implies that the predicted steepening of the SB profiles with $\ionfrac$ (see Sec.~\ref{sec:reionhistory} below) and velocity off-set are (conservatively) underestimated by our calculations.

We point out that constraints on the reionization history that are provided by current observations of the Ly$\alpha$ forest and the cosmic microwave background, as well as theoretical investigations, suggest that  the ionization fraction at $z=7.7$ was mostly complete (i.e. well in excess of $\ionfrac \sim 0.50$; see e.g. Fig.~11 of \citealt{Pritchard10} or Fig.~4 of \citealt{ciardi11}). This suggest that for realistic constraints on $\ionfrac$ at $\zz=7.7$ the scatter beyond the cube is not significant, and so we do not include this extra scatter in the rest of our analysis.

\subsubsection{Clustering of Nearby Sources}
\label{sec:cluster}

While a roughly uniform ionization fraction could be induced by an ionizing background, photons from neighboring haloes clustered around a source would induce further fluctuations in  the gas ionization structure. To investigate the impact this has on our reference simulation, we propagate ionizing photons from all the stellar sources present in the cube (referred to as 'case $\mathcal C$' for clustering), rather than only from the central source. Figure~\ref{fig:clusterion} shows a slice through the ionization structure obtained in case $\mathcal C$ (middle panel) along with that from the reference simulation (left panel). As expected, the ionization bubble around the object is much larger and complex when the effect of all the sources is considered. In case $\mathcal C$, the final value for the volume averaged ionization fraction is $\ionfrac=0.340$, compared to 0.047 of the reference case.

To understand how the ionization structure due to clustered sources affects the propagation of Ly$\alpha$ photons compared to a uniform ionization, we do another test run (referred to as 'case $\mathcal U$' for uniform) producing the same final volume averaged ionization fraction of case $\mathcal C$, i.e. 0.340, but this is induced by photons emitted only from the central source embedded in a gas with a uniform initial ionization fraction of 0.293. Case $\mathcal U$ is shown in the right panel of Figure~\ref{fig:clusterion} and exhibits a very different ionization structure compared to case $\mathcal C$. While the region with fully ionized gas is smaller in case $\mathcal U$, the situation is reversed in the region beyond the bubble with some fully neutral gas still present in case $\mathcal C$. Note that in all cases the Ly$\alpha$ photons are emitted only by the central source. 

Figure~\ref{fig:clustersb} shows the SB maps for the three cases discussed above (for the top right panel in Fig.~\ref{fig:run1sb}). Both case $\mathcal C$ and case $\mathcal U$ show a more concentrated SB profile compared to the reference run as expected for a lower $\ionfrac$, but the details in the SB maps differ due to the distribution of neutral gas. In particular, all the maps (from the six faces of the cube) of case $\mathcal U$ have a more concentrated structure compared to case $\mathcal C$ due to the higher ionization level of the gas outside the fully ionized bubble. In fact, since the ionization bubble is not large enough for the Ly$\alpha$ photons to be redshifted out of resonance when they reach its border, the neutral gas distribution beyond it is particularly important for the determination of the corresponding SB maps. Because the concentration in the SB improves detectability, case $\mathcal U$ is expected to favor observations of LAEs. At $SB_{\rm th} = 10^{-21}$~erg~s$^{-1}$~cm$^{-2}$~arcsec$^{-2}$, case $\mathcal C$ has $\fesclyaigm^{\rm eff}\sim0.001$, while case $\mathcal U$ has $\sim0.004$, compared to the reference case of $\sim 0.0007$. 

It might seem that clustering of neighbouring sources does not substantially change the observability of the LAE, but because the flux is redistributed in different directions due to the different ionization structure in the IGM, $\fesclyaigm^{\rm eff}$ can vary by orders of magnitude depending on the observing direction. For the same $SB_{\rm th}$, another direction (top right panel in Fig.~\ref{fig:run1sb}) has $\fesclyaigm^{\rm eff}=1.5$, 0.7 and 0.6 for case $\mathcal C$, case $\mathcal U$ and the reference case, respectively. For reference, the input luminosity is $L_{\rm Ly\alpha}^{\rm esc} = 9.2 \times 10^{41}$ erg~s$^{-1}$. Thus it should be kept in mind that even though both neighbouring sources and uniform background flux improves detectability of LAEs by increasing the mean ionization fraction of IGM in the region, the detailed ionization structure around a source plays a crucial role in determining the SB profiles in different directions. 

\section{Effects on Estimates of the Reionization History}
\label{sec:reionhistory}

\begin{figure*}
\centering
\includegraphics[width=130mm,height=130mm]{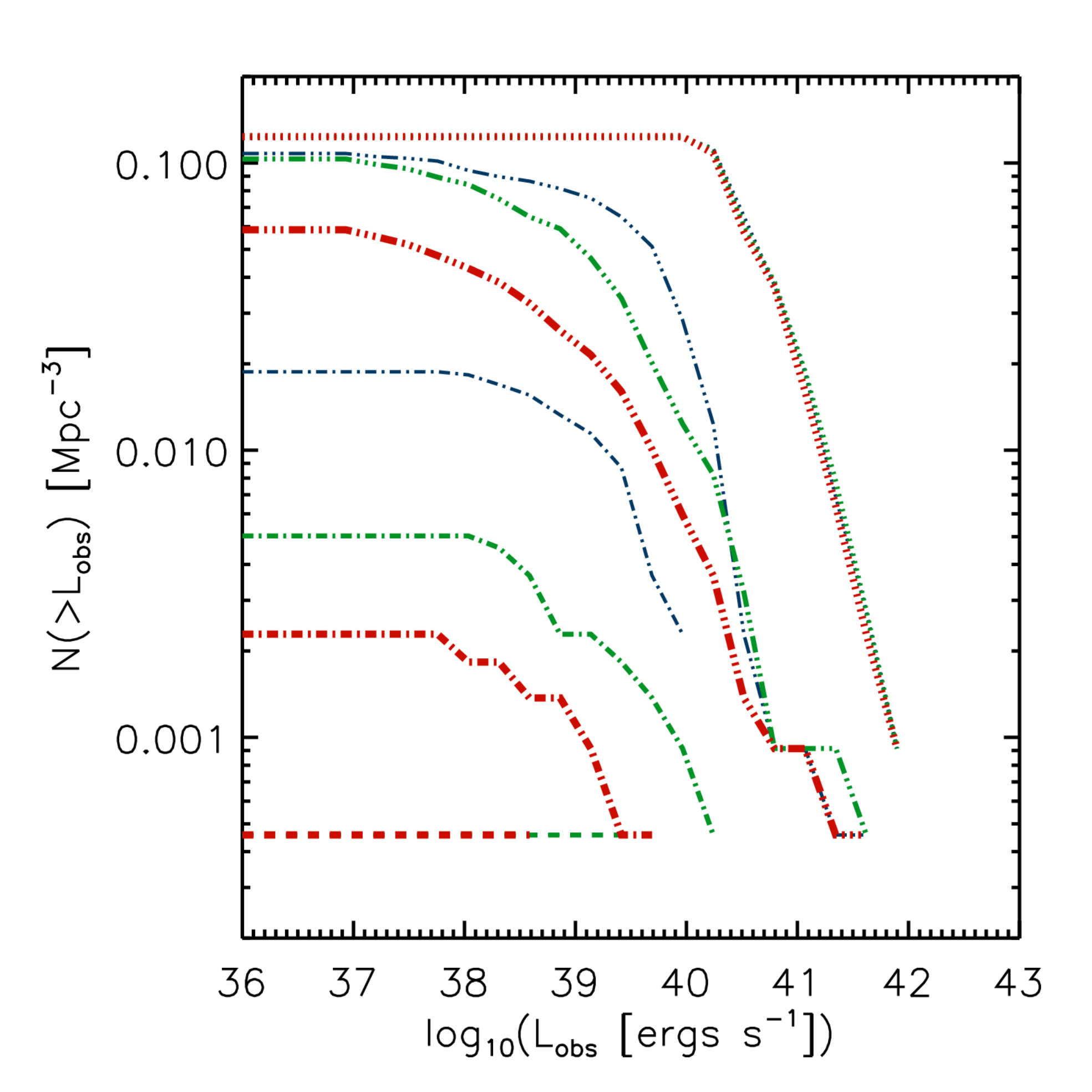}
\caption[Luminosity functions.]{Luminosity functions calculated for the 45 most massive objects in L05 for different SB cuts. The runs are for initial uniform volume averaged ionization fraction $\ionfrac$=0 (thick red), $\ionfrac=0.5$ (medium thick green) and $\ionfrac=0.89$ (thin blue). The SB thresholds are  $SB_{\rm th}=10^{-25}$ erg~s$^{-1}$~cm$^{-2}$~arcsec$^{-2}$ (dotted lines), $10^{-21}$ erg~s$^{-1}$~cm$^{-2}$~arcsec$^{-2}$ (triple dot dashed lines), $10^{-20}$ erg~s$^{-1}$~cm$^{-2}$~arcsec$^{-2}$ (dot dashed lines) and $5 \times 10^{-20}$ erg~s$^{-1}$~cm$^{-2}$~arcsec$^{-2}$ (dashed lines).}
\label{fig:lumfunc}
\end{figure*}

Keeping in mind the discussions in the previous Sections, we now turn to study how structure in the IGM affects the methods which use the SB profiles of LAEs to estimate the mean ionization fraction of the universe. One of the most popular methods is to calculate the luminosity function of LAEs and determine its redshift evolution \citep[e.g.][]{malhotra04,dayal09,kashikawa11}. Using analytic models to describe the reionization history, the above authors estimate the evolution of $\tau$ (employing equations similar to Eq.~\ref{eqn:tau2}). The intrinsic LF of LAEs is then calculated from halo mass functions, assuming that the luminosity of an object is proportional to its mass and that it is attenuated by ${\rm e}^{-\tau}$ due IGM absorption/scattering. A comparison with the observed LF provides a constraint on the actual $\tau$ and thus on the gas neutral fraction in the IGM. But these methods rely on the assumption that observations of LAEs obtain the total flux from the source detected as a point source. Some recent observations though have detected faint extended SB profiles \citep[][]{steidel11}, as produced in our simulations. As we have seen, the probability of detection of a LAE strongly depends on $SB_{\rm th}$, especially for low $\ionfrac$. Here we investigate how a different choice of $SB_{\rm th}$ affects the LAE LFs for different values of $\ionfrac$, aiming at a qualitative assessment, while a more detailed and quantitative comparison to observations is part of a future study. 

We model the 45 most massive objects in L05 using $\ionfrac$=0, 0.5 and 0.89 as described in Section~\ref{sec:unifion}. For each object, we calculate the six SB profiles and their respective observed luminosities $L_{\rm obs} (> SB_{\rm th})$, which yields a total of 270 data points (LAEs) at each $SB_{\rm th} = 5 \times 10^{-20}, ~10^{-20}, ~10^{-21}$ and $10^{-25}$~erg~s$^{-1}$~cm$^{-2}$~arcsec$^{-2}$. We consider the six sides of a simulation cube as part of six separate observational fields with comoving side of $5$~$h^{-1}$ Mpc each. Structure in the IGM leads to differences in the SB profiles for the six sides (as seen in Fig.~\ref{fig:run1sb}), which in turn leads to a scatter in the observed LFs calculated from the different fields. Therefore, LFs are calculated separately for each of the observational fields and then averaged in each luminosity bin to obtain the mean luminosity functions, which are shown in Figure~\ref{fig:lumfunc}.

First we focus on $\ionfrac=0$, to understand the effects of SB thresholds on detections (as seen in Figure~\ref{fig:lumdmsb}), and determine the luminosity functions.  At $SB_{\rm th} = 5 \times 10^{-20}$~erg~s$^{-1}$~cm$^{-2}$~arcsec$^{-2}$ only one object is detected with an observed luminosity of $L_{\rm obs} (> SB_{\rm th}) = 3 \times 10^{38}$ erg~s$^{-1}$, while $SB_{\rm th} = 10^{-20}$~erg~s$^{-1}$~cm$^{-2}$~arcsec$^{-2}$ leads to the detection of 5 LAEs with observed luminosities larger than in the previous case. The LF is thus shifted to higher values in both axis. Going one order of magnitude deeper in SB leads to a huge increase in the number of detections ($\sim 130$ LAEs). This gives rise to a shift in the LF of about two orders of magnitude in both luminosity and number density. At $SB_{\rm th} = 10^{-25}$~erg~s$^{-1}$~cm$^{-2}$~arcsec$^{-2}$ all the flux in all the LAEs is observed, returning the intrinsic LF of the simulated LAE sample. It should be kept in mind that the value of $SB_{\rm th}$ needed to observe the full flux depends on the intrinsic luminosity of the sources and neutral IGM structure around it.

\begin{figure*}\centering\includegraphics[width=140mm,height=100mm]{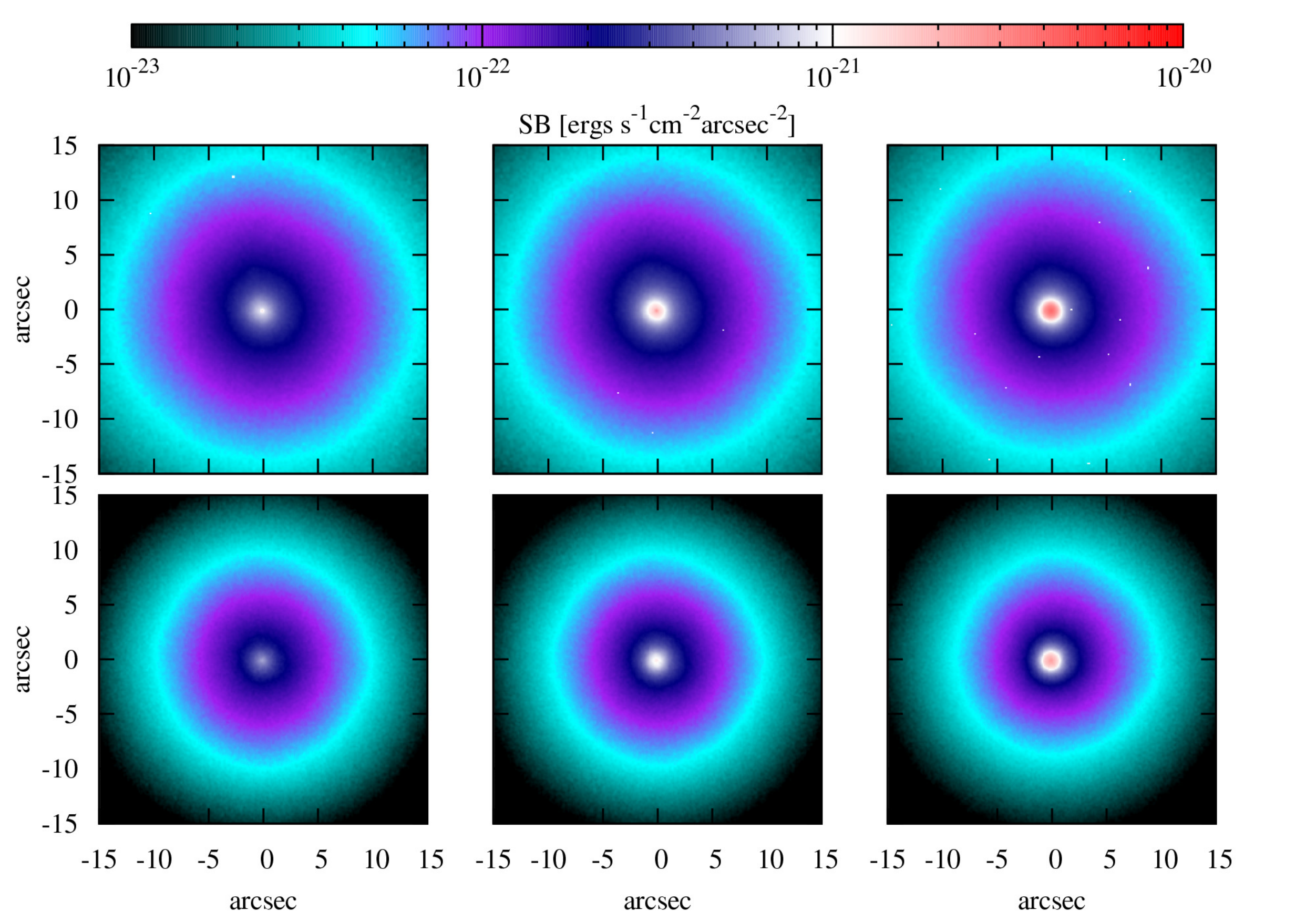}
\caption[Stacked SB profiles.]{Stacked SB profiles for initial volume averaged ionization fractions of $\ionfrac=$0 (left panels), 0.5 (middle) and 0.89 (right). The top (bottom) panels refer to the mean (median) values in each pixel of the stacked SB profiles.}
\label{fig:sbstack}
\end{figure*}

In the real universe, both the mean ionization fraction and the intrinsic properties of galaxies change. This leads to a complex evolution of the LF of LAEs with redshift. In addition, the different $SB_{\rm th}$ from real observations add another level of complexity to the redshift evolution of observed LFs. To help in constraining $\ionfrac$ with observations of LFs, here we try to understand how the LF depends on the ionization fraction of the IGM for different choices of $SB_{\rm th}$.  For $SB_{\rm th} = 10^{-25}$~erg~s$^{-1}$~cm$^{-2}$~arcsec$^{-2}$, the luminosity functions for different values of $\ionfrac$ are equivalent, because all the flux from the sources is detected. For lower SB cuts instead, the number of detections changes with $\ionfrac$, leading to an increasing number density of LAEs at a given luminosity for higher $\ionfrac$. The difference between LFs at different $\ionfrac$ is higher for smaller values of $SB_{\rm th}$. It should be noted that shallow detections (e.g. $10^{-20}$~erg~s$^{-1}$~cm$^{-2}$~arcsec$^{-2}$) at higher $\ionfrac$ (e.g. $\ionfrac=0.89$) have similar LFs as deeper but incomplete detections  (e.g. $10^{-21}$~erg~s$^{-1}$~cm$^{-2}$~arcsec$^{-2}$) at lower $\ionfrac$ (e.g. $\ionfrac=0$). Thus, when comparing luminosity functions at different redshifts, the SB threshold of the observations should be taken into account to obtain a realistic  estimate of the decrease in the number density of LAEs. Another point to note is that discrepancies between LFs for different values of $SB_{\rm th}$ are smaller at higher ionization fractions (e.g. $\ionfrac=0.89$). Thus only mild differences in LFs for orders of magnitude in $SB_{\rm th}$ at a specific redshift can be used as an evidence for high $\ionfrac$. At high $\ionfrac$, absorption outside the edge of the cube (as discussed in Section~\ref{sec:unifion}), which has not been included in this test, would lead to an even larger gap between LFs for different $SB_{\rm th}$. Nevertheless one has to keep in mind that velocity shifts discussed in Section~\ref{sec:velshift}  (e.g. due to outflows from the galaxy) could also lead to milder changes in luminosity functions with changing $SB_{\rm th}$. 

Due to the difficulties in obtaining deep observations of LAEs (refer to Section~\ref{sec:results}), we look into the possibility of gathering additional information from stacked profiles. This was recently done by \cite{steidel11} at $\zz=3$ to estimate the low SB extended emission present in individual non-detections. In addition, also \cite{zheng10b} looked at estimating total Ly$\alpha$ luminosities from stacked profiles for different mass and luminosity bins. In our study, we investigate the shape of the stacked SB profiles without binning for halo properties, to understand how the profiles are changed by $\ionfrac$. The data set used is the same one discussed above. In general, for $\ionfrac=0$ the individual profiles show a flatter SB profile with substantial structure due to scattering through a mostly neutral but structured IGM. For higher initial ionization fractions, the strength of the point source increases as more photons escape with less scattering resulting in a steeper SB profile. These characteristics can be exploited to study the mean ionization fraction of the universe from stacked LAE SB profiles.

Figure~\ref{fig:sbstack} shows the stacked SB maps from the 270 simulated maps of the 45 galaxies described previously. The top (bottom) panels refer to the mean (median) values in each pixel of the stack. The mean value is generally affected by the outliers when the distribution is not gaussian. In our stacks, the higher mass haloes act as outliers being brighter and more point-source-like due to larger ionized bubbles around them, but are fewer in number compared to the majority of SB maps of low mass haloes. This leads to the mean profile being steeper and brighter than the median profile at each $\ionfrac$. But it is important to note that in both mean and median stacks, the trend of the stacked SB profiles with $\ionfrac$ is the same.

As we have seen in Section~\ref{sec:unifion}, the higher $\ionfrac$ is, the steeper is the SB profile of a LAE. This trend is more important for haloes of lower masses, as the ionization bubble around these objects are smaller due to the lower ionizing photon flux, leading to a more diffuse SB profile at low $\ionfrac$. As $\ionfrac$ increases, a larger fraction of the photons from these haloes escapes unscattered, leading to a more concentrated SB profile. For high mass haloes, even at low $\ionfrac=0$, the photons escape with lesser scatter due to larger ionized regions around them, leading to a steeper SB profile. Due to the shape of the halo mass function though, there is a much larger number of low mass haloes in a comoving volume of the universe. Therefore, when stacking SB profiles of LAEs in a volume, the average of the stack is dominated by the low mass haloes. Thus the stacked SB profile at low $\ionfrac=0$ is flat. The stacked profile has no structural details, because the profiles in individual SB maps have a random distribution, thus averaging out in the stack. At ionization fractions of $\ionfrac=0.5,~0.89$, the photons escape easier in all directions leading to a steeper SB profile in all directions. This also appears when stacking SB profiles, with the stacked SB profile becoming more concentrated with increasing $\ionfrac$. Thus this method of stacking can be used to estimate the steepness of the profile and in turn the volume averaged ionization fraction of the universe at that redshift.  As already discussed, the velocity shifts could also lead to the steepening of the individual SB profile even at $\ionfrac=0$, thus reducing the difference in concentration between stacked profiles for different $\ionfrac$. But absorption due to IGM outside the cube at $\ionfrac=0$ would lead to flattening of individual SB profiles. This would help in differentiating between $\ionfrac$ in stacked SB profiles.

In short, a lot of information about the galaxies and their surrounding IGM can be obtained from the deep observations of LAE SB profiles. A more detailed modelling and comparison with observations is the work of a future study.

\section{Discussion and conclusions}
\label{sec:discussion}

The Epoch of Reionization (EoR) is an interesting and important event in the history of the universe, the time being when H in the gas changed from a mainly neutral to a highly ionized state. The details of the process are still unclear and are the topic of a large number of theoretical and observational studies. Lyman Alpha Emitters (LAEs), due to their strong Ly$\alpha$ emission line which scatters for even tiny amounts of neutral hydrogen, are used as one of the tools to probe the EoR. Modelling LAEs is a complex task with a large number of parameters and scales being important. Different studies either focus on a detailed understanding of individual aspects of LAEs or use numerical/semi-analytic approaches to model the LAEs and compare them to observations. In this paper, we focus on the Inter-Galactic Medium (IGM)  close to the ionizing/Ly$\alpha$ source and its effect on the observability of LAEs due to Ly$\alpha$ RT. The IGM close to the source is important: the density, temperature and ionization structure controls the Ly$\alpha$ radiative transfer (RT), determining the intensity and surface brightness (SB) distribution along different lines-of-sight.

To achieve this, we simulated a sample of more than $100$ LAEs, using galaxies and surrounding IGM extracted from simulation boxes of 5-30$~\Mpch$ at $\zz=7.7$. Coupled RT of ionizing and Ly$\alpha$ photons was performed using CRASH$\alpha$, to determine the ionization structure carved in the IGM by the ionizing photons exiting from the galaxies and the spectrum of the Ly$\alpha$ photons scattering through the remaining neutral hydrogen along different lines-of-sight. The outputs were also used to produce Ly$\alpha$ SB maps for the lines-of-sight perpendicular to the six sides of each simulation cube. Analyses were done to study individual objects as well as statistical trends. A parameter study was also undertaken to understand the dependence of the results on the different, currently uncertain factors involved in the simulations.

Our main results are:
\begin{enumerate}
\item Inhomogeneities in the IGM affects Ly$\alpha$ RT, leading to structure in the simulated SB maps of LAEs. The anisotropic escape of Ly$\alpha$ photons through the IGM causes large variations in the total flux in SB maps along different lines-of-sight for the same object. This leads to $\sim 30\%$ more scatter in the observed luminosity-mass relation than the intrinsic one for our sample of simulated LAEs. This also leads to a spread in the values for the escape fraction of Ly$\alpha$ photons from the IGM: $\fesclyaigm=0.73\pm0.18~(1\sigma)$. Note that some lines-of-sight, especially through voids, can have $\fesclyaigm>1$ due to Ly$\alpha$ photon contributions from other lines-of-sight by scattering and higher probability of Ly$\alpha$ photon escape towards the observer through voids.

\item Observational campaigns have SB thresholds, $SB_{\rm th}$, which limit the fraction of the flux observed from the SB distribution. At $SB_{\rm th}$ (e.g.  $~ 10^{-20}$ erg~s$^{-1}$~cm$^{-2}$~arcsec$^{-2}$), the observability of the LAE strongly depends on the IGM ionization structure and velocity field around the objects. Therefore, in observational campaigns at very low $SB_{\rm th}$, a single detection within the observed comoving volume needs not be that of the most massive LAE in the region but could be a lower mass object seen through a biased line-of-sight. At even fainter $SB_{\rm th}$, more detections are possible, but the observed luminosity values can vary by orders of magnitude for a single object depending on the line-of-sight. To obtain the total flux from the object along each line-of-sight, one needs unrealistically deep observations at $SB_{\rm th} \sim 10^{-25}$ erg~s$^{-1}$~cm$^{-2}$~arcsec$^{-2}$. Thus we find that the impact of SB thresholds on estimates of observed luminosity is very important and needs to be taken into account in the calculation of luminosity functions of LAEs from observational campaigns (also see Zheng et al. 2010a).

\item One of the main factors which affect the Ly$\alpha$ RT through the IGM is the wavelength of the Ly$\alpha$ photons when they escape from the Inter-Stellar Medium (ISM) into the IGM. Outflows/winds in the ISM would redshift the photons reducing their probability of scattering in the IGM and aiding the escape of Ly$\alpha$ photons towards the observer \citep{dijkstra10}. We find that the higher the redshifting of the Ly$\alpha$ photons before entering the IGM is, the more concentrated is the SB profile, making the detection easier at low $SB_{\rm th}$ and improving the observability of LAEs. Therefore, outflows from the galaxies could weaken the imprint of reionization on the statistics of LAEs than thought previously. But due to the uncertainty in the link between the galaxy and outflows, this would add a challenge to estimating the ionization fraction of the IGM at high redshifts. 

\item The ionization structure of the IGM also plays a very important role in Ly$\alpha$ RT thus determining the SB profiles. We estimate the effects of an ionizing background using RT simulations with an initial non-zero uniform ionization level.  At low levels of mean ionization in the IGM ($\ionfrac\sim 0$), the Ly$\alpha$ photons undergo scattered diffusion through the IGM and the SB profiles thus produced are more extended and faint,  making it harder to observe the total flux from the object. At high levels of mean IGM ionization ($\ionfrac >  0.5$), the photons scatter less, leading to a more point source-like SB profile, making it easier to detect more flux for low $SB_{\rm th}$. Clustering of sources also affects the ionization structure around the object by making the region more ionized, but with a complex structure for the ionized bubble due to the distribution of neighbouring ionizing sources and IGM gas distribution. The qualitative effect on the SB profiles due to clustering is similar to that of a uniform ionization case with the same mean ionization level. But the detailed distribution of flux along different lines-of-sight varies due to the differences in the IGM ionization structure. Thus, this shows that proper treatment of the ionization structure in the region is important for better modelling LAEs.

\item Due to the difficulty in achieving very high $SB_{\rm th}$ in observational campaigns, stacking of SB maps can be used to extract more information from the current LAE samples mapped upto relatively lower $SB_{\rm th}$ \citep[e.g.][]{steidel11,zheng10b}. We find that the mean/median stacked SB profile of LAEs becomes steeper at higher mean IGM ionization levels thus giving an additional technique to use LAEs to study EoR. But outflow and halo properties among others could also lead to steepening of SB profiles and should to be considered when using this technique.
\end{enumerate}

\section*{Acknowledgments}
Many thanks to the anonymous referee for helpful comments, to Rudiger Pakmor for providing us the GadgettoGrid code, to Ouchi Masami for sensitivity estimates of LAEs' surveys and to Andrea Ferrara for stimulating comments. AJD acknowledges support from and participation in the International Max Planck Research School on Astrophysics at the Ludwig-Maximilians University. AM acknowledges the support of the DFG Priority Program 1177.

\label{lastpage}


\begin{thebibliography}{MBNB03}

\bibitem[\protect\citeauthoryear{Adams}{1975}]{adams75} Adams 
T.~F., 1975, ApJ, 201, 350 
\bibitem[\protect\citeauthoryear{Barkana 
\& Loeb}{2001}]{barkana01} Barkana R., Loeb A., 2001, PhR, 349, 125
\bibitem[\protect\citeauthoryear{Barnes et al.}{2011}]{barnes11} 
Barnes L.~A., Haehnelt M.~G., Tescari E., Viel M., 2011, MNRAS, 416, 1723 
\bibitem[\protect\citeauthoryear{Bland-Hawthorn 
\& Putman}{2001}]{bland01} Bland-Hawthorn J., Putman M.~E., 2001, ASPC, 240, 369 
\bibitem[\protect\citeauthoryear{Bolton et al.}{2011}]{bolton11} 
Bolton J.~S., Haehnelt M.~G., Warren S.~J., Hewett P.~C., Mortlock D.~J., 
Venemans B.~P., McMahon R.~G., Simpson C., 2011, MNRAS, 416, L70 
\bibitem[\protect\citeauthoryear{Bromm 
\& Loeb}{2003}]{bromm03} Bromm V., Loeb A., 2003, Nature, 425, 812 
\bibitem[\protect\citeauthoryear{Cantalupo et 
al.}{2005}]{cantalupo05} Cantalupo S., Porciani C., Lilly S.~J., 
Miniati F., 2005, ApJ, 628, 61 
\bibitem[\protect\citeauthoryear{Cassata et 
al.}{2011}]{cassata11} Cassata P., et al., 2011, A\&A, 525, A143 
\bibitem[\protect\citeauthoryear{Ciardi et al.}{2001}]{ciardi01} 
Ciardi B., Ferrara A., Marri S., Raimondo G., 2001, MNRAS, 324, 381 
\bibitem[\protect\citeauthoryear{Ciardi, Bianchi, 
\& Ferrara}{2002}]{ciardi02} Ciardi B., Bianchi S., Ferrara A., 2002, MNRAS, 331, 463 
\bibitem[\protect\citeauthoryear{Ciardi et al.}{2011}]{ciardi11} 
Ciardi B., Bolton J.~S., Maselli A., Graziani L., 2011, arXiv, 
arXiv:1112.4646 
\bibitem[\protect\citeauthoryear{Charlot 
\& Fall}{1993}]{charlot93} Charlot S., Fall S.~M., 1993, ApJ, 415, 580 
\bibitem[\protect\citeauthoryear{Clarke 
\& Oey}{2002}]{clarke02} Clarke C., Oey M.~S., 2002, MNRAS, 337, 1299 
\bibitem[\protect\citeauthoryear{Cl{\'e}ment et 
al.}{2012}]{clement11} Cl{\'e}ment B., et al., 2012, A\&A, 538, A66 
\bibitem[\protect\citeauthoryear{Dav{\'e}, Finlator, 
\& Oppenheimer}{2006}]{dave06} Dav{\'e} R., Finlator K., Oppenheimer B.~D., 2006, MNRAS, 370, 273 
\bibitem[\protect\citeauthoryear{Davis 
\& Wilkinson}{1974}]{davis74} Davis M., Wilkinson D.~T., 1974, ApJ, 192, 251 
\bibitem[\protect\citeauthoryear{Davis et al.}{1985}]{davis85} 
Davis M., Efstathiou G., Frenk C.~S., White S.~D.~M., 1985, ApJ, 292,
371
\bibitem[\protect\citeauthoryear{Dawson et al.}{2004}]{dawson04} 
Dawson S., Malhotra S., Rhoads J.~E., Stern D., Dey A., Spinrad H., Jannuzi 
B.~T., 2004, AAS, 36, 746 
\bibitem[\protect\citeauthoryear{Dayal et al.}{2009}]{dayal09} 
Dayal P., Ferrara A., Saro A., Salvaterra R., Borgani S., Tornatore L., 
2009, MNRAS, 400, 2000 
\bibitem[\protect\citeauthoryear{Dijkstra, Lidz, 
\& Wyithe}{2007}]{dijkstra07a} Dijkstra M., Lidz A., Wyithe J.~S.~B., 2007, MNRAS, 377, 1175 
\bibitem[\protect\citeauthoryear{Dijkstra, Wyithe, 
\& Haiman}{2007}]{dijkstra07b} Dijkstra M., Wyithe J.~S.~B., Haiman Z., 2007, MNRAS, 379, 253 
\bibitem[\protect\citeauthoryear{Dijkstra 
\& Wyithe}{2007}]{dijkstra07c} Dijkstra M., Wyithe J.~S.~B., 2007, MNRAS, 379, 1589 
\bibitem[\protect\citeauthoryear{Dijkstra 
\& Loeb}{2008}]{dijkstra08} Dijkstra M., Loeb A., 2008, MNRAS, 391, 457 
\bibitem[\protect\citeauthoryear{Dijkstra 
\& Wyithe}{2010}]{dijkstra10} Dijkstra M., Wyithe J.~S.~B., 2010, MNRAS, 408, 35
\bibitem[\protect\citeauthoryear{Dijkstra, Mesinger, 
\& Wyithe}{2011}]{dijkstra11} Dijkstra M., Mesinger A., Wyithe J.~S.~B., 2011, MNRAS, 414, 2139 
\bibitem[Dijkstra 
\& Hultman Kramer(2012)]{dijkstra12} Dijkstra, M., \& Hultman Kramer, R.\ 2012, submitted to MNRAS, arXiv:1203.3803 
\bibitem[\protect\citeauthoryear{Djorgovski 
\& Thompson}{1992}]{djorgovski92} Djorgovski S., Thompson D.~J., 1992, IAUS, 149, 337 
\bibitem[\protect\citeauthoryear{Dove, Shull, 
\& Ferrara}{2000}]{dove00} Dove J.~B., Shull J.~M., Ferrara A., 2000, ApJ, 531, 846 
\bibitem[\protect\citeauthoryear{Eisenstein et 
al.}{2005}]{eisenstein05} Eisenstein D.~J., et al., 2005, ApJ, 633, 
560 
\bibitem[\protect\citeauthoryear{Forero-Romero et 
al.}{2011}]{forero11} Forero-Romero J.~E., Yepes G., 
Gottl{\"o}ber S., Knollmann S.~R., Cuesta A.~J., Prada F., 2011, MNRAS, 
415, 3666 
\bibitem[\protect\citeauthoryear{Fujita et al.}{2003}]{fujita03} 
Fujita A., Martin C.~L., Mac Low M.-M., Abel T., 2003, ApJ, 599, 50 
\bibitem[\protect\citeauthoryear{Furlanetto, Zaldarriaga, 
\& Hernquist}{2006}]{furlanetto06} Furlanetto S.~R., Zaldarriaga M., Hernquist L., 2006, MNRAS, 365, 1012 
\bibitem[\protect\citeauthoryear{Gnedin, Kravtsov, 
\& Chen}{2008}]{gnedin08} Gnedin N.~Y., Kravtsov A.~V., Chen H.-W., 2008, ApJ, 672, 765 
\bibitem[\protect\citeauthoryear{Gronwall et 
al.}{2011}]{gronwall10} Gronwall C., Bond N.~A., Ciardullo R., 
Gawiser E., Altmann M., Blanc G.~A., Feldmeier J.~J., 2011, ApJ, 743, 9 
\bibitem[\protect\citeauthoryear{Guaita et al.}{2010}]{guaita10} 
Guaita L., et al., 2010, ApJ, 714, 255 
\bibitem[\protect\citeauthoryear{Gunn 
\& Peterson}{1965}]{gunn65} Gunn J.~E., Peterson B.~A., 1965, ApJ, 142, 1633 
\bibitem[\protect\citeauthoryear{Haiman 
\& Spaans}{1999}]{haiman99} Haiman Z., Spaans M., 1999, ApJ, 518, 138 
\bibitem[\protect\citeauthoryear{Haiman}{2002}]{haiman02} Haiman 
Z., 2002, ApJ, 576, L1 
\bibitem[\protect\citeauthoryear{Hartmann et 
al.}{1988}]{hartmann88} Hartmann L.~W., Huchra J.~P., Geller 
M.~J., O'Brien P., Wilson R., 1988, ApJ, 326, 101 
\bibitem[\protect\citeauthoryear{Hill et al.}{2008}]{hill08} 
Hill G.~J., et al., 2008, ASPC, 399, 115 
\bibitem[\protect\citeauthoryear{Hu 
\& McMahon}{1996}]{hu96} Hu E.~M., McMahon R.~G., 1996, Natur, 382, 231 
\bibitem[\protect\citeauthoryear{Iliev et al.}{2008}]{iliev08} 
Iliev I.~T., Shapiro P.~R., McDonald P., Mellema G., Pen U.-L., 2008, 
MNRAS, 391, 63 
\bibitem[\protect\citeauthoryear{Iwata et al.}{2009}]{iwata09} 
Iwata I., et al., 2009, ApJ, 692, 1287 
\bibitem[\protect\citeauthoryear{Iye et al.}{2006}]{iye06} 
Iye M., et al., 2006, Natur, 443, 186 
\bibitem[\protect\citeauthoryear{Kashikawa et 
al.}{2011}]{kashikawa11} Kashikawa N., et al., 2011, ApJ, 734, 119 
\bibitem[\protect\citeauthoryear{Kobayashi, Kamaya, 
\& Yonehara}{2006}]{kobayashi06} Kobayashi M.~A.~R., Kamaya H., Yonehara A., 2006, ApJ, 636, 1 
\bibitem[\protect\citeauthoryear{Kobayashi, Totani, 
\& Nagashima}{2007}]{kobayashi07} Kobayashi M.~A.~R., Totani T., Nagashima M., 2007, ApJ, 670, 919 
\bibitem[\protect\citeauthoryear{Kollmeier et 
al.}{2010}]{kollmeier10} Kollmeier J.~A., Zheng Z., Dav{\'e} R., 
Gould A., Katz N., Miralda-Escud{\'e} J., Weinberg D.~H., 2010, ApJ, 708, 
1048 
\bibitem[\protect\citeauthoryear{Laursen 
\& Sommer-Larsen}{2007}]{laursen07} Laursen P., Sommer-Larsen J., 2007, ApJ, 657, L69 
\bibitem[\protect\citeauthoryear{Laursen, Sommer-Larsen, 
\& Andersen}{2009}]{laursen09} Laursen P., Sommer-Larsen J., Andersen A.~C., 2009, ApJ, 704, 1640 
\bibitem[\protect\citeauthoryear{Laursen, Sommer-Larsen, 
\& Razoumov}{2011}]{laursen11} Laursen P., Sommer-Larsen J., Razoumov A.~O., 2011, ApJ, 728, 52 
\bibitem[\protect\citeauthoryear{Le Delliou et 
al.}{2005}]{ledelliou05} Le Delliou M., Lacey C., Baugh C.~M., 
Guiderdoni B., Bacon R., Courtois H., Sousbie T., Morris S.~L., 2005, 
MNRAS, 357, L11 
\bibitem[\protect\citeauthoryear{Le Delliou et 
al.}{2006}]{ledelliou06} Le Delliou M., Lacey C.~G., Baugh C.~M., 
Morris S.~L., 2006, MNRAS, 365, 712 
\bibitem[\protect\citeauthoryear{Lehnert et 
al.}{2010}]{lehnert10} Lehnert M.~D., et al., 2010, Natur, 467, 
940 
\bibitem[\protect\citeauthoryear{Leitherer et 
al.}{1999}]{leitherer99} Leitherer C., et al., 1999, ApJS, 123, 3 
\bibitem[\protect\citeauthoryear{Loeb 
\& Rybicki}{1999}]{loeb99} Loeb A., Rybicki G.~B., 1999, ApJ, 524, 527 
\bibitem[\protect\citeauthoryear{Maio et al.}{2007}]{maio07} 
Maio U., Dolag K., Ciardi B., Tornatore L., 2007, MNRAS, 379, 963 
\bibitem[\protect\citeauthoryear{Maio et 
al.}{2009}]{maio09} Maio U., Ciardi B., Yoshida N., Dolag K., Tornatore L., 2009, A\&A, 503, 25 
\bibitem[\protect\citeauthoryear{Maio et al.}{2010}]{maio10} 
Maio U., Ciardi B., Dolag K., Tornatore L., Khochfar S., 2010, MNRAS, 407, 
1003 
\bibitem[\protect\citeauthoryear{Malhotra 
\& Rhoads}{2002}]{malhotra02} Malhotra S., Rhoads J.~E., 2002, ApJ, 565, L71 
\bibitem[\protect\citeauthoryear{Malhotra 
\& Rhoads}{2004}]{malhotra04} Malhotra S., Rhoads J.~E., 2004, ApJ, 617, L5 
\bibitem[\protect\citeauthoryear{Malhotra 
\& Rhoads}{2006}]{malhotra06} Malhotra S., Rhoads J.~E., 2006, ApJ, 647, L95 
\bibitem[\protect\citeauthoryear{Maselli, Ferrara, 
\& Ciardi}{2003}]{maselli03} Maselli A., Ferrara A., Ciardi B., 2003, MNRAS, 345, 379 
\bibitem[\protect\citeauthoryear{Maselli, Ciardi, 
\& Kanekar}{2009}]{maselli09} Maselli A., Ciardi B., Kanekar A., 2009, MNRAS, 393, 171 
\bibitem[\protect\citeauthoryear{McQuinn et 
al.}{2007}]{mcquinn07} McQuinn M., Hernquist L., Zaldarriaga M., 
Dutta S., 2007, MNRAS, 381, 75 
\bibitem[\protect\citeauthoryear{Meier 
\& Terlevich}{1981}]{meier81} Meier D.~L., Terlevich R., 1981, ApJ, 246, L109 
\bibitem[\protect\citeauthoryear{Mesinger 
\& Furlanetto}{2008a}]{mesinger08a} Mesinger A., Furlanetto S.~R., 2008a, MNRAS, 386, 1990 
\bibitem[\protect\citeauthoryear{Mesinger 
\& Furlanetto}{2008b}]{mesinger08b} Mesinger A., Furlanetto S.~R., 2008b, MNRAS, 385, 1348 
\bibitem[\protect\citeauthoryear{Miralda-Escude 
\& Rees}{1998}]{miralda98} Miralda-Escude J., Rees M.~J., 1998, ApJ, 497, 21 
\bibitem[\protect\citeauthoryear{Nagamine et 
al.}{2010}]{nagamine10} Nagamine K., Ouchi M., Springel V., 
Hernquist L., 2010, PASJ, 62, 1455 
\bibitem[\protect\citeauthoryear{Neufeld}{1991}]{neufeld91} 
Neufeld D.~A., 1991, ApJ, 370, L85 
\bibitem[\protect\citeauthoryear{Nilsson et 
al.}{2011}]{nilsson11} Nilsson K.~K., {\"O}stlin G., M{\o}ller P., M{\"o}ller-Nilsson O., Tapken C., Freudling W., Fynbo J.~P.~U., 2011, A\&A, 529, A9 
\bibitem[\protect\citeauthoryear{Nilsson et 
al.}{2007}]{nilsson07} Nilsson K.~K., et al., 2007, A\&A, 471, 71 
\bibitem[\protect\citeauthoryear{Ono et al.}{2012}]{ono11} 
Ono Y., et al., 2012, ApJ, 744, 83 
\bibitem[\protect\citeauthoryear{Ouchi et al.}{2008}]{ouchi08} 
Ouchi M., et al., 2008, ApJS, 176, 301 
\bibitem[\protect\citeauthoryear{Ouchi et al.}{2009}]{ouchi09} 
Ouchi M., et al., 2009, ApJ, 696, 1164 
\bibitem[\protect\citeauthoryear{Ouchi et al.}{2010}]{ouchi10} 
Ouchi M., et al., 2010, ApJ, 723, 869 
\bibitem[\protect\citeauthoryear{Paardekooper et 
al.}{2011}]{paardekooper11} Paardekooper J.-P., Pelupessy F.~I., Altay G., Kruip C.~J.~H., 2011, A\&A, 530, A87 
\bibitem[\protect\citeauthoryear{Pakmor}{2010}]{pakmor10} 
Pakmor R., 2010, PhD thesis, TU Munich.
\bibitem[\protect\citeauthoryear{Partridge}{1974}]{partridge74} 
Partridge R.~B., 1974, ApJ, 192, 241 
\bibitem[\protect\citeauthoryear{Partridge 
\& Peebles}{1967}]{partridge67} Partridge R.~B., Peebles P.~J.~E., 1967, ApJ, 147, 868 
\bibitem[Pentericci et al.(2011)]{pentericci11} Pentericci, L., 
Fontana, A., Vanzella, E., et al.\ 2011, ApJ, 743, 132 
\bibitem[\protect\citeauthoryear{Pierleoni, Maselli, 
\& Ciardi}{2009}]{pierleoni09} Pierleoni M., Maselli A., Ciardi B., 2009, MNRAS, 393, 872 
\bibitem[Pritchard et al.(2010)]{Pritchard10} Pritchard, J.~R., 
Loeb, A., \& Wyithe, J.~S.~B.\ 2010, MNRAS, 408, 57 
\bibitem[\protect\citeauthoryear{Razoumov 
\& Sommer-Larsen}{2006}]{razoumov06} Razoumov A.~O., Sommer-Larsen J., 2006, ApJ, 651, L89 
\bibitem[\protect\citeauthoryear{Rhoads 
\& Malhotra}{2001}]{rhoads01} Rhoads J.~E., Malhotra S., 2001, ApJ, 563, L5 
\bibitem[\protect\citeauthoryear{Rhoads et al.}{2004}]{rhoads04} 
Rhoads J.~E., et al., 2004, ApJ, 611, 59 
\bibitem[\protect\citeauthoryear{Ricotti 
\& Shull}{2000}]{ricotti00} Ricotti M., Shull J.~M., 2000, ApJ, 542, 548 
\bibitem[\protect\citeauthoryear{Santoro 
\& Shull}{2006}]{santoro06} Santoro F., Shull J.~M., 2006, ApJ, 643, 26 
\bibitem[\protect\citeauthoryear{Santos}{2004}]{santos04} Santos 
M.~R., 2004, MNRAS, 349, 1137 
\bibitem[\protect\citeauthoryear{Samui, Srianand, 
\& Subramanian}{2009}]{samui09} Samui S., Srianand R., Subramanian K., 2009, MNRAS, 398, 2061 
\bibitem[\protect\citeauthoryear{Schaerer}{2003}]{schaerer03} Schaerer D., 2003, A\&A, 397, 527 
\bibitem[\protect\citeauthoryear{Schaerer et 
al.}{2011}]{schaerer11} Schaerer D., Hayes M., Verhamme A., Teyssier R., 2011, A\&A, 531, A12 
\bibitem[\protect\citeauthoryear{Schenker et 
al.}{2012}]{schenker11} Schenker M.~A., Stark D.~P., Ellis R.~S., 
Robertson B.~E., Dunlop J.~S., McLure R.~J., Kneib J.-P., Richard J., 2012, 
ApJ, 744, 179 
\bibitem[\protect\citeauthoryear{Schneider et 
al.}{2003}]{schneider03} Schneider R., Ferrara A., Salvaterra R., 
Omukai K., Bromm V., 2003, Natur, 422, 869 
\bibitem[\protect\citeauthoryear{Schneider et 
al.}{2006}]{schneider06} Schneider R., Omukai K., Inoue A.~K., 
Ferrara A., 2006, MNRAS, 369, 1437 
\bibitem[\protect\citeauthoryear{Shapley et 
al.}{2006}]{shapley06} Shapley A.~E., Steidel C.~C., Pettini M., 
Adelberger K.~L., Erb D.~K., 2006, ApJ, 651, 688 
\bibitem[\protect\citeauthoryear{Siana et al.}{2010}]{siana10} 
Siana B., et al., 2010, ApJ, 723, 241 
\bibitem[\protect\citeauthoryear{Spitzer}{1978}]{spitzer78} 
Spitzer L., 1978, ppim.book,  
\bibitem[\protect\citeauthoryear{Springel}{2005}]{springel05} 
Springel V., 2005, MNRAS, 364, 1105
\bibitem[\protect\citeauthoryear{Steidel, Pettini, 
\& Adelberger}{2001}]{steidel01} Steidel C.~C., Pettini M., Adelberger K.~L., 2001, ApJ, 546, 665 
\bibitem[\protect\citeauthoryear{Steidel et 
al.}{2011}]{steidel11} Steidel C.~C., Bogosavljevi{\'c} M., 
Shapley A.~E., Kollmeier J.~A., Reddy N.~A., Erb D.~K., Pettini M., 2011, 
arXiv, arXiv:1101.2204 
\bibitem[\protect\citeauthoryear{Taniguchi et 
al.}{2005}]{taniguchi05} Taniguchi Y., et al., 2005, PASJ, 57, 165 
\bibitem[\protect\citeauthoryear{Tasitsiomi}{2006}]{tasitsiomi06} 
Tasitsiomi A., 2006, ApJ, 645, 792 
\bibitem[\protect\citeauthoryear{Tilvi et al.}{2009}]{tilvi09} 
Tilvi V., Malhotra S., Rhoads J.~E., Scannapieco E., Thacker R.~J., Iliev 
I.~T., Mellema G., 2009, ApJ, 704, 724 
\bibitem[\protect\citeauthoryear{Tilvi et al.}{2010}]{tilvi10} 
Tilvi V., et al., 2010, ApJ, 721, 1853 
\bibitem[\protect\citeauthoryear{Tolman}{1934}]{tolman34} Tolman 
R.~C., 1934, rtc..book,  
\bibitem[\protect\citeauthoryear{Tornatore et 
al.}{2007}]{tornatore07} Tornatore L., Borgani S., Dolag K., 
Matteucci F., 2007, MNRAS, 382, 1050 
\bibitem[\protect\citeauthoryear{Trenti et al.}{2010}]{trenti10} 
Trenti M., Smith B.~D., Hallman E.~J., Skillman S.~W., Shull J.~M., 2010, 
ApJ, 711, 1198 
\bibitem[\protect\citeauthoryear{Valls-Gabaud}{1993}]{valls93} 
Valls-Gabaud D., 1993, ApJ, 419, 7 
\bibitem[Verhamme et 
al.(2008)]{verhamme07} Verhamme, A., Schaerer, D., Atek, H., \& Tapken, C.\ 2008, A\&A, 491, 89 
\bibitem[\protect\citeauthoryear{Verhamme, Schaerer, 
\& Maselli}{2006}]{verhamme06} Verhamme A., Schaerer D., Maselli A., 2006, A\&A, 460, 397 
\bibitem[\protect\citeauthoryear{Wang et al.}{2009}]{wang09} 
Wang J.-X., Malhotra S., Rhoads J.~E., Zhang H.-T., Finkelstein S.~L., 
2009, ApJ, 706, 762 
\bibitem[\protect\citeauthoryear{Wise 
\& Cen}{2009}]{wise09} Wise J.~H., Cen R., 2009, ApJ, 693, 984 
\bibitem[\protect\citeauthoryear{Wolf et al.}{2010}]{wolf10} 
Wolf C., et al., 2010, AAS, 42, \#410.10 
\bibitem[\protect\citeauthoryear{Wood 
\& Loeb}{2000}]{wood00} Wood K., Loeb A., 2000, ApJ, 545, 86 
\bibitem[\protect\citeauthoryear{Wyithe 
\& Loeb}{2005}]{wyithe05} Wyithe J.~S.~B., Loeb A., 2005, ApJ, 625, 1 
\bibitem[\protect\citeauthoryear{Yajima et al.}{2009}]{yajima09} 
Yajima H., Umemura M., Mori M., Nakamoto T., 2009, MNRAS, 398, 715 
\bibitem[\protect\citeauthoryear{Yajima, Choi, 
\& Nagamine}{2011}]{yajima11a} Yajima H., Choi J.-H., Nagamine K., 2011, MNRAS, 412, 411 
\bibitem[\protect\citeauthoryear{Yajima et al.}{2011}]{yajima11b} 
Yajima H., Li Y., Zhu Q., Abel T., 2011, arXiv, arXiv:1109.4891 
\bibitem[\protect\citeauthoryear{Yoshida et 
al.}{2003}]{yoshida03} Yoshida N., Abel T., Hernquist L., 
Sugiyama N., 2003, ApJ, 592, 645 
\bibitem[\protect\citeauthoryear{Zheng et al.}{2010}]{zheng10} 
Zheng Z., Cen R., Trac H., Miralda-Escud{\'e} J., 2010, ApJ, 716, 574 
\bibitem[Zheng et al.(2011)]{zheng10b} Zheng, Z., Cen, R., 
Weinberg, D., Trac, H., \& Miralda-Escud{\'e}, J.\ 2011, ApJ, 739, 62 

\end{thebibliography}
\end{document}